\documentclass[journal,comsoc]{IEEEtran}
\usepackage{algorithm}
\usepackage{algpseudocode}
\usepackage{amsfonts}
\usepackage{bm}
\usepackage{amsmath}
\usepackage{cite}
\usepackage{amssymb}
\usepackage{enumerate}
\usepackage{color,xcolor}
\usepackage{graphicx}
\usepackage{multirow,tabularx}
\usepackage{subfigure}
\usepackage{hyperref}
\usepackage[T1]{fontenc}

\makeatletter

\newcommand{\Rmnum}[1]{\expandafter\@slowromancap\romannumeral #1@}
\makeatother

\graphicspath{{figure/}}

\newcommand{\BE}{\begin{equation}}
\newcommand{\EE}{\end{equation}}
\newcommand{\BS}{\begin{subequations}}
\newcommand{\ES}{\end{subequations}}
\renewcommand{\bf}{\bm}

\usepackage{stfloats}
\ifCLASSINFOpdf

\else

\fi

\begin{document}

\title{{Gaussian Message Passing for Overloaded  \\ Massive MIMO-NOMA}}

\author{\IEEEauthorblockN{Lei Liu, \emph{Member, IEEE,} Chau Yuen, \emph{Senior Member, IEEE,} Yong Liang Guan, \emph{Senior Member, IEEE,}\\ Ying Li, \emph{Member, IEEE,} and Chongwen Huang, \emph{Student Member, IEEE}}
%\thanks{This work was supported in part by the National Natural Science Foundation of China under Grants 61671345 and 61750110529, and joint fund of ministry of education of People's Republic of China 6141A02022338. The first author was also supported by the China Scholarship Council under Grant 20140690045. The editor coordinating the review of this paper and approving it for publication was Lars Rasmussen. (Corresponding author: Ying Li.)}
\thanks{Lei Liu was with the State Key Lab of Integrated Services Networks, Xidian University, Xi'an, 710071, China. He is now with the Department of Electronic Engineering, City University of Hong Kong, Hong Kong, SAR, China. (e-mail: lliu\_0@stu.xidian.edu.cn)}
\thanks{Chau Yuen and Chongwen Huang are with the Singapore University of Technology and Design, Singapore (e-mail: yuenchau@sutd.edu.sg, chongwen\_huang@mymail.sutd.edu.sg)}
\thanks{Yong Liang Guan is with the School of Electrical and Electronic Engineering, Nanyang Technological University, Singapore (e-mail: eylguan@ntu.edu.sg).}
\thanks{Ying Li is with the State Key Lab of Integrated Services Networks, Xidian University, Xi'an, 710071, China (e-mail: yli@mail.xidian.edu.cn).}
\thanks{This work has been presented in part at the 2016 IEEE GLOBECOM, Washington, DC USA \cite{lei_GC16}.}
}

%\thanks{Lei Liu was with the State Key Lab of Integrated Services Networks, Xidian University, Xi'an, 710071, China. He is now with the Department of Electronic Engineering, City University of Hong Kong, Hong Kong, SAR, China. (e-mail: lliu\_0@stu.xidian.edu.cn). Chau Yuen and Chongwen Huang are with the Singapore University of Technology and Design, Singapore (e-mail: yuenchau@sutd.edu.sg, chongwen\_huang@mymail.sutd.edu.sg). Yong Liang Guan is with the School of Electrical and Electronic Engineering, Nanyang Technological University, Singapore (e-mail: eylguan@ntu.edu.sg). Ying Li is with the State Key Lab of Integrated Services Networks, Xidian University, Xi'an, 710071, China (e-mail: yli@mail.xidian.edu.cn).}
%
%}

\markboth{IEEE Transactions on Wireless Communications,~Vol.~XX, No.~X, XXX~2018}%
{Shell \MakeLowercase{\textit{et al.}}: Bare Demo of IEEEtran.cls for Journals}

\maketitle

\begin{abstract}
This paper considers a low-complexity Gaussian Message Passing (GMP) scheme for a coded massive Multiple-Input Multiple-Output (MIMO) systems with Non-Orthogonal Multiple Access (massive MIMO-NOMA), in which a base station with $N_s$ antennas serves $N_u$ sources simultaneously in the same frequency. Both $N_u$ and $N_s$ are large numbers, and we consider the overloaded cases with $N_u>N_s$. The GMP for MIMO-NOMA is a message passing algorithm operating on a fully-connected loopy factor graph, which is well understood to fail to converge due to the correlation problem. In this paper, we utilize the large-scale property of the system to simplify the convergence analysis of the GMP under the overloaded condition. First, we prove that the \emph{variances} of the GMP definitely converge to the mean square error (MSE) of Linear Minimum Mean Square Error (LMMSE) multi-user detection. Secondly, the \emph{means} of the traditional GMP will fail to converge when $ N_u/N_s< (\sqrt{2}-1)^{-2}\approx5.83$. Therefore, we propose and derive a new convergent GMP called scale-and-add GMP (SA-GMP), which always converges to the LMMSE multi-user detection performance for any $N_u/N_s>1$, and show that it has a faster convergence speed than the traditional GMP with the same complexity. Finally, numerical results are provided to verify the validity and accuracy of the theoretical results presented.
\end{abstract}

\begin{IEEEkeywords}
Overloaded massive MIMO-NOMA, convergence improvement, Gaussian message passing, loopy factor graph, low-complexity detection.
\end{IEEEkeywords}

\IEEEpeerreviewmaketitle
\section{Introduction}
{Recently, massive \emph{Multiple-Input and Multiple-Output} (MIMO), in which the Base Station (BS) has a large number of antennas (e.g., hundreds or even more), has attracted more and more attention \cite{Rusek2013, 5GWhitepaper, Marzetta2010 , METIS, Ngo2012, Dai2013}. In particular, massive MIMO is able to bring significant improvement both in throughput and energy efficiency \cite{Rusek2013, Ngo2012, Dai2013, 5GWhitepaper, Marzetta2010 , METIS}. Furthermore, the access demand of wireless communication has increased exponentially in these years. Based on the forecast from ABI Research, the number of wireless communication devices will reach 40.9 billion in 2020 \cite{ABI}, and the units of Internet of Things (IoT) are predicted to increase to 26 billion by 2020 \cite{Gartner}. As a result, due to the limited spectrum resources, overloaded access in the same time/frequency/code is inevitable in the future wireless communication systems \cite{Lien}. \emph{Non-Orthogonal Multiple Access} (NOMA), in which all the users can be served concurrently at the same time/frequency/code, has been identified as one of the key radio access technologies to increase the spectral efficiency and reduce the latency of the Fifth Generation (5G) mobile networks \cite{Wang_NOMA1,Dai_NOMA,Benjebbour_NOMA,Imari_NOMA,GL2018,Di2018,LYW2017}. In addition, the system performance of NOMA can be further enhanced by combining NOMA with MIMO (MIMO-NOMA)\cite{LiPing2017,YC2018,Jiang2018, Ding2016,Kim2015,Wang_NOMA, Lei20161}.

In this paper, an overloaded massive MIMO-NOMA uplink, in which the number of users $N_u$ is larger than the number of antennas $N_s$, is considered, i.e., $N_u>N_s$. It is unlike the conventional massive MIMO which requires $N_u<<N_s$ \cite{Marzetta2010,Rusek2013}. We shall propose a low-complexity iterative detection for the overloaded massive MIMO-NOMA. Furthermore, the convergence of the iterative detection is analyzed.
\begin{figure*}[t]
  \centering
  \includegraphics[width=18.0cm]{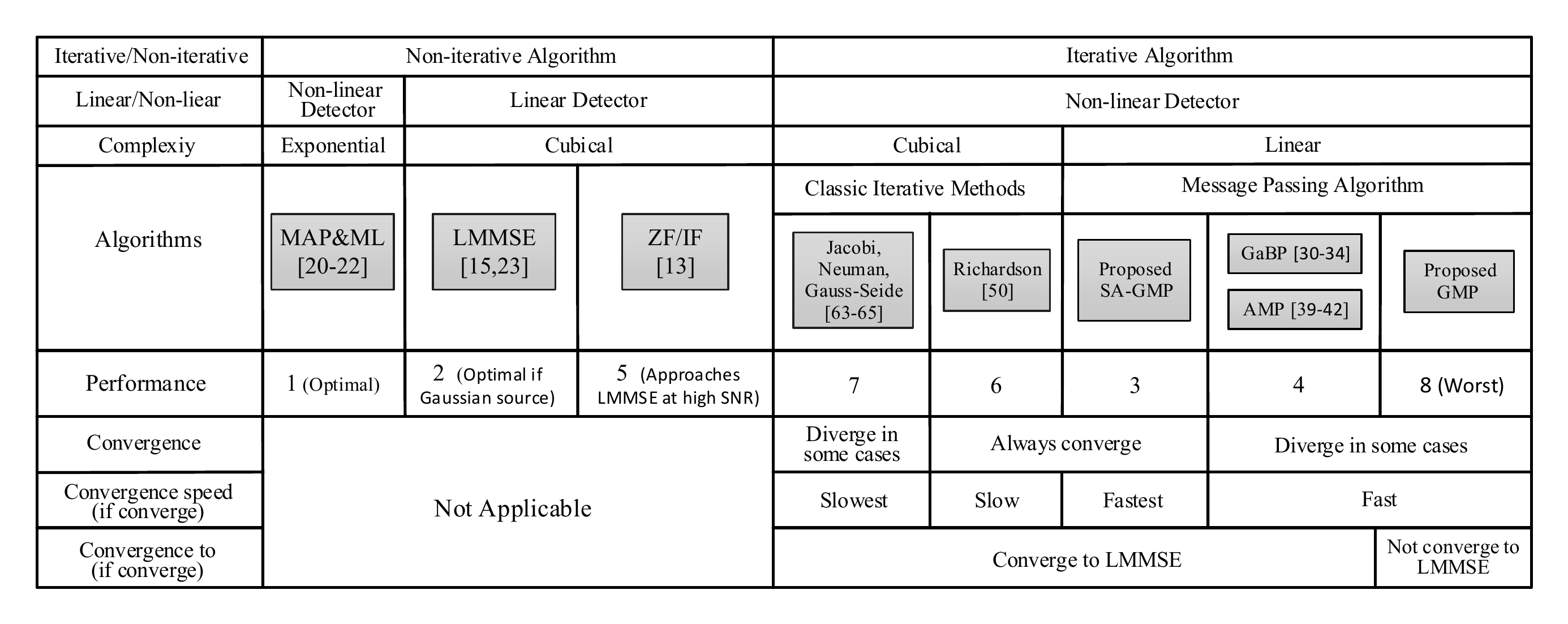}\\\vspace{-0.2cm}
  \caption{An overview of algorithms for the overloaded MIMO-NOMA system. }\label{connect}
\end{figure*}
\subsection{Motivation and Related Work}
Unlike the \emph{Orthogonal Multiple Access} (OMA) systems, \emph{e.g.} the \emph{Time Division Multiple Access }(TDMA) and \emph{Orthogonal Frequency Division Multiple Access} (OFDMA) \emph{etc.} \cite{Liao_OMA,Bolcskei_OMA}, the signal processing for the NOMA will impose higher computational complexity and energy consumption at the \emph{base station} (BS)\cite{Dai_NOMA,Benjebbour_NOMA,Imari_NOMA,LYW2017}. Low-complexity uplink detection for MIMO-NOMA is a challenging problem due to the non-orthogonal interference between the users \cite{Kim2015,Lei20161,YC2018}, especially when the number of users is large. For example, the optimal \emph{Multiuser Detector} (MUD) for the \emph{Multiuser MIMO} (MU-MIMO), such as the \emph{Maximum A-posteriori Probability} (MAP) detection or \emph{Maximum Likelihood} (ML) detection, was proven to be NP-hard \cite{Micciancio2001,verdu1984_1,verdu1983}. Low-complexity uplink detection for MIMO-NOMA is hence desirable \cite{Kim2015, YC2018, Lei20161}. In \cite{Lei20161}, a low-complexity iterative \emph{Linear Minimum Mean Square Error} (LMMSE) detection was proposed to approach the sum capacity of the MIMO-NOMA system, but the complexity of LMMSE detection is still too high due to the need of performing large matrix inversion \cite{tse2005}. To avoid the matrix inversion, a graph based detection called \emph{Message Passing Algorithm} (MPA) is applied, which is encountered in many computer science and engineering problems such as signal processing, linear programming, social networks, etc. \cite{Forney2001, kschischang2001, Loeliger2004, Loeliger2006, Guo2008, William2009}.
There are different types of MPAs for MUD, such as belief propagation on Gaussian graphical model (Ga-BP) and Gaussian belief propagation (GaBP) algorithms \cite{Weiss20012, Roy2001, malioutov2006, Su2014, Su2015}, BP on ``non-Gaussian" probabilistic models \cite{andrea2005,yoon2014}. These BP algorithms exchange extrinsic information in the process. Please note the different meanings of ``Gaussian" in GaBP and Ga-BP: ``Gaussian" in GaBP means that the messages in BP are Gaussian (represented by mean and variance), while ``Gaussian" in Ga-BP means that the system can be described by a Gaussian probabilistic model. Since Gaussian graphical problems can be efficiently solved by GaBP, Ga-BP is presumed to be GaBP by default. Separately, the authors in \cite{Lei2016} look into another class of MPA for MUD: Gaussian Message Passing (GMP) that exchanges extrinsic and/or \emph{a-posteriori} information. In addition, SA-GMP is proposed to improve the MSE performance and convergence properties via spectral radius minimization \cite{Lei2016}.

Apart from that, in \cite{Donoho2009,Donoho2010a,Donoho2011,Bayati2011}, an advanced Approximate Message Passing (AMP), which is asymptotically optimal under many practical scenarios, is proposed. AMP is simple to implement in practice, i.e., its complexity is as low as $O(N_sN_u)$ per iteration. However, AMP may become unreliable in ill conditioned channels \cite{Samuel2017, Ma2017}.

For factor graphs with a tree structure, the means and variances of the MPA converge to the true marginal means, if LLRs are used as messages and the sum-product algorithm is applied \cite{kschischang2001, Loeliger2004}. However, if the graph has cycles (loops), the MPA may fail to converge. In general, on loopy graphs, different types of MPAs may have different convergence behaviors, such as different convergence conditions/speeds/performances. To the best of our knowledge, most previous works of the MPA focus on the convergence of GaBP algorithms for Gaussian graphs \cite{Weiss20012,malioutov2006,Su2014, Su2015}. {In \cite{Roy2001, andrea2005}, the convergence of message passing algorithms are analyzed. However, the messages in \cite{andrea2005} are for vector variables, which need high-complexity matrix operations during the message updating \cite{Roy2001}. Therefore, the result in \cite{andrea2005} is only applied for Code Division Multiple Access (CDMA) MIMO with binary channels. For the case of underloaded massive MIMO-NOMA, in \cite{Lei2016} the authors have analyzed the convergence of} {underloaded GMP and underloaded SA-GMP, which are distinct from the overloaded GMP and overloaded SA-GMP in this paper}. However, the convergence of GMP for the overloaded massive MIMO-NOMA is far from solved. In Fig. \ref{connect}, we give an overview of the detection algorithms for the overloaded massive MIMO-NOMA system.
}

\subsection{Contributions}
In this paper, we first analyse the convergence of GMP for the overloaded massive MIMO-NOMA system. We discover that the convergence of GMP depends on {a} spectral radius $\rho_{\tiny GMP}$, i.e., the estimation $\hat{\mathbf{x}}(\tau)$ of GMP at the $\tau$th iteration converges to a fixed point ${\mathbf{x}}^*$ with a square error ${\|\hat{\mathbf{x}}(\tau)-{\mathbf{x}}^*\|_2}\varpropto\rho_{\tiny GMP}^{\tau}$. Specifically, the GMP converges when {{the spectral radius}} is less than 1, i.e., $\rho_{\tiny GMP}<1$, otherwise it diverges.
Furthermore, we optimize {{the spectral radius}} with linear modifications (e.g. scale and add), and obtain a new low-complexity fast-convergence MUD for coded overloaded ($\beta  = {N_u \mathord{\left/{\vphantom {N_u N_s}} \right. \kern - \nulldelimiterspace} N_s}>1$) massive MIMO-NOMA system. The contributions of this paper are summarized below.

 \noindent
\hangafter=1
\setlength{\hangindent}{1.4em} $\bullet$ We prove that the variances of GMP definitely converge to the MSE of LMMSE detection. This provides an alternative way to estimate the MSE of the LMMSE detector.

 \noindent
\hangafter=1
\setlength{\hangindent}{1.4em} $\bullet$ We prove that the convergence of GMP depends on {{a spectral radius}}.

 \noindent
\hangafter=1
\setlength{\hangindent}{1.4em} $\bullet$ Two sufficient conditions for which the means of GMP converge to a higher MSE than those of the LMMSE detector for $\{\beta: \beta>(\sqrt{2}-1)^{-2}\}$ are derived.

 \noindent
\hangafter=1
\setlength{\hangindent}{1.4em} $\bullet$ A new fast-convergence detector called  scale-and-add GMP (SA-GMP), which converges to the LMMSE detection, and has a faster convergence speed than GMP for any $\{\beta: \beta>1\}$, is proposed. %This detector minimizes the spectral radius.
 %\newpage
\subsection{Comparisons with literature}
Here we compare our contribution with the literatures:
{\subsubsection{Relationship with AMP and damping AMP}
The asymptotic variance behavior of AMP is the same as GMP \cite{Cakmak2015, Meng2015}, because for large MIMO systems, the approximation in AMP is accurate, which is guaranteed by the \emph{Law of Large Numbers} (LLN) \cite{Donoho2010a,Bayati2011}. However, their means behave differently because GMP exchanges APP information at the sum node, while AMP is derived by extrinsic-information-based message passing. As a result, the mean of GMP may have a worse MSE than AMP/GaBP due to the correlation problem in APP processing. However, the proposed optimized SA-GMP has better MSE and convergence properties than GMP and AMP/GaBP. Different from AMP (whose each step can be rigorously described by state evolution (SE)), SA-GMP only requires its convergence properties (e.g. the final fixed point, convergence condition and speed) to be optimized. Furthermore, although the damping schemes \cite{Vila2015,Parker2014} have been used to prevent the divergence of AMP, there are fundamental differences between the damping AMP and SA-GMP. First, in the damping AMP, the damping operation is employed in every message update step, including mean update and variance update. However, in SA-GMP, the scale-and-add (SA) modification is only applied in the mean message update at the sum node, while the variable-node message update and the variance update remain unchanged from the GMP. Second, in the damping AMP, there is no closed-form solution for the optimal damping parameter, which can only be determined empirically, or by exhaustive search. In contrast, the optimal relaxation parameter in SA-GMP has been mathematically derived in a closed form by minimizing the spectral radius (see Corollary 3).
}

\subsubsection{Differences from the underloaded case} Due to the overloaded number of users, the expression of the GMP is different from the underloaded case in \cite{ Lei2016}. Hence, the {underloaded GMP and underloaded SA-GMP} have poor performance when applied in overloaded scenario \cite{Lei2016}. Furthermore, the convergence results of GMP for the overloaded massive MIMO-NOMA are also different from that in \cite{Lei2016}: \emph{i)} the convergence condition is different; \emph{ii)} if the GMP for the overloaded massive MIMO-NOMA converges, it converges to a higher (rather than the same) MSE than that of the LMMSE detector.

This paper is organized as follows. In Section II, the overloaded massive MIMO-NOMA and LMMSE estimator are introduced. The GMP is elaborated in Section III. Section IV presents the proposed fast-convergence SA-GMP. Numerical results are shown in Section V, and we conclude our work in Section VI.

\section{System Model and LMMSE Estimator}
\begin{figure}[ht!]
  \centering
  \includegraphics[width=9cm]{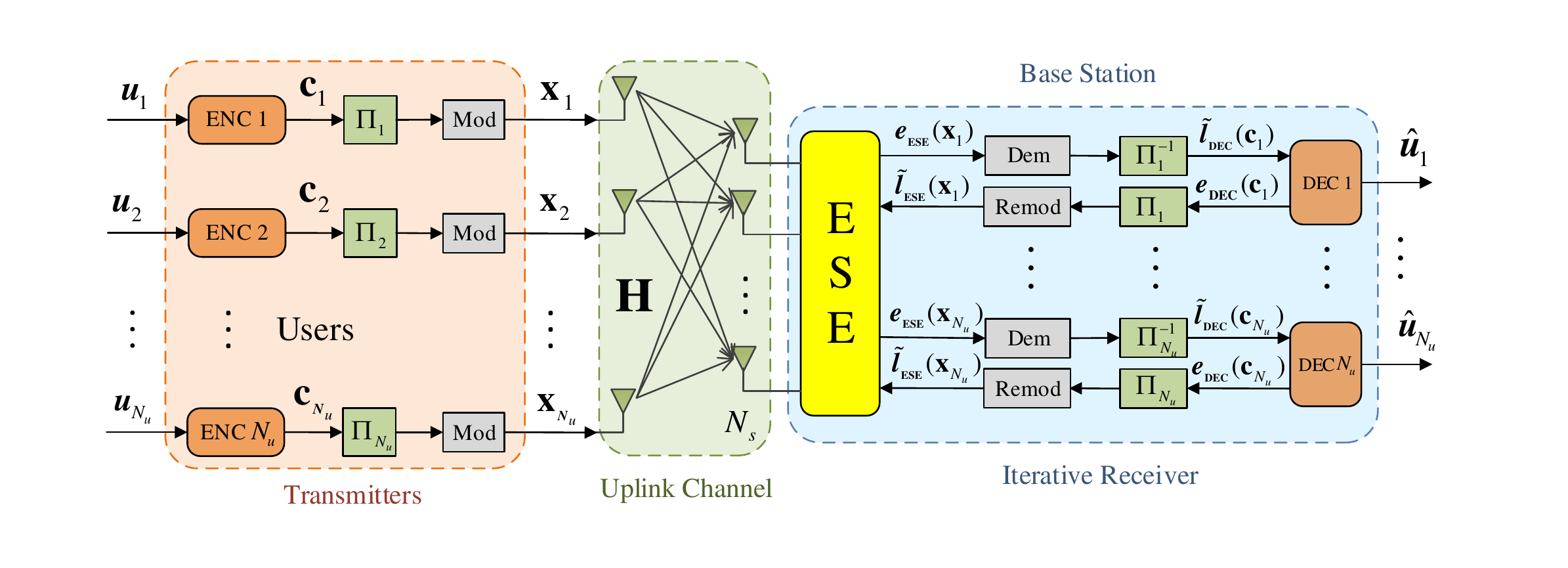}\\
  \caption{Coded overloaded massive MIMO-NOMA system: $N_{u}$ autonomous single-antenna terminals communicate with an array of $N_s$ antennas at the BS in the same time and frequency, where $N_u>N_s$.  The BS estimates the sources by an iterative receiver from the received signals. ENC denotes the error-correcting encoder and DEC denotes the corresponding decoder. {``Remod" and ``Dem" denote the re-modulator and demodulator respectively.} $\Pi_i$ denotes interleaver and $\Pi_i^{-1}$ denotes de-interleaver. ESE represents the elementary signal estimator, which is basically a MIMO multi-user detector. $\mathbf{H}$ is the small-scale fading channel matrix.}\label{f2}
\end{figure}
{Fig. \ref{f2} shows a coded overloaded massive MIMO-NOMA system, in which $N_{u}$ autonomous single-antenna terminals simultaneously communicate with the BS which has an array of $N_s$ antennas \cite{Marzetta2010,Rusek2013}. Both $N_{u}$ and $N_s$ are large numbers, and $N_{u}>N_s$. The $N_s\times1$ received ${ \mathbf{y}_t}$ at time $t$ is \vspace{-0.1cm}
\begin{equation}\label{e1}
{ \mathbf{y}_t}= \mathbf{H}{ \mathbf{x}^{tr}_t} + \mathbf{n}_t,\quad t\in \mathcal{N},\vspace{-0.1cm}
\end{equation}
where $\mathcal{N}=\{1,\cdots,N\}$, $\mathbf{H}\sim\mathcal{N}^{N_s\times N_u}(0,1)$ is a fading channel matrix, $\mathbf{n}_t\sim\mathcal{N}^{{N_s}}(0,\sigma_n^2)$ is an $N_s \times 1$ Gaussian noise vector, and $\mathbf{x}^{tr}_t=[x_{1,t},\cdots,x_{N_u,t}]^T$ is the message vector sent from $N_{u}$ users at time $t$. We assume that the BS knows the $\mathbf{H}$, and only the real system is considered because the complex case can be easily extended from the real case\footnote{A complex system model ${ \tilde{\mathbf{y}}_t}= \tilde{\mathbf{H}}{ \tilde{\mathbf{x}}^{tr}_t} + \tilde{\mathbf{n}}_t$
can be converted to a corresponding real-valued system model \eqref{e1},
where $\mathbf{y}_t=[\Re\{\tilde{\mathbf{y}}_t\}, \Im\{\{\tilde{\mathbf{y}}_t\}\}]^T$ is the $2N_s\times 1$ real-valued signal vector, accordingly ${\mathbf{x}}^{tr}_t=[\Re\{\tilde{\mathbf{x}}^{tr}_t\}, \Im\{\tilde{\mathbf{x}}^{tr}_t\}]^T$, ${\mathbf{n}}_t=[\Re\{\tilde{\mathbf{n}}_t\}, \Im\{\tilde{\mathbf{n}}_t\}]^T$ and
${\mathbf{H}} = \left[ {\begin{array}{*{20}{c}}
{\Re\{\tilde{\mathbf{H}}\}}&{-\Im\{\tilde{\mathbf{H}}\}}\\
{\Im\{\tilde{\mathbf{H}}\}}&{\Re\{\tilde{\mathbf{H}}\}}
\end{array}} \right].$ Note that $\Re\{\cdot\}$ and $\Im\{\cdot\}$ denote the real part and imaginary part respectively.} \cite{Gao2014}. Let $\hat{\mathbf{x}}_t$ be the estimated vector of ${\mathbf{x}^{tr}_t}$, the mean square error (MSE) of the estimation is defined as $MSE=\frac{1}{N_u}\|\hat{\mathbf{x}}_t-{\mathbf{x}}^{tr}_t\|_2^2$.

\subsection{Transmitters}
{As illustrated in Fig. \ref{f2}, at user $i$, an information sequence ${\bf{u}}_i \in\{0,1\}^K$ is encoded by a common (used by all users) error-correcting code (with code rate $\frac{K}{N_c}$) into an $N_c$-length coded sequence ${\mathbf{c}}_{i}$, $\mathop{i}\in \mathcal{N}_{u},\; \mathcal{N}_{u}= \left\{ {{{1,2,}} \cdots {{,N_{u}}}} \right\}$, which is interleaved by an {$N_c$-length} independent random interleaver $\Pi_{i}$ and then is modulated by a Gaussian modulator to obtain a transmission vector $\mathbf{x}_{i}=[x_{i,1},x_{i,2},\cdots,x_{i,N}]^T$.} In this paper, we assume that each $\mathbf{x}_{i}$ is independent identically Gaussian distributed, i.e., $\mathbf{x}_{i}\sim \mathcal{N}^N(0,\sigma^2_{x_i})$ for $i\in \mathcal{N}_u$. The source variance $\sigma^2_{x_i}$ denotes the power constraint or the large-scale fading coefficient of each user. To simplify the analysis, we assume $\sigma^2_{x_i}=1$ for $\forall i\in \mathcal{N}_u$.

Regarding the Gaussian transmission assumption, in the real communication system, discrete modulated signals are generally used. However, according to the Shannon theory \cite{Cover2006,Gamal2012}, the capacity of a Gaussian channel is achieved by a Gaussian input. Therefore, the independent Gaussian source assumption is widely used in the design of communication networks \cite{ Kafedziski2012,  Liu2012, Matamoros2011}. In practice, the capacity-achieving SCM, the quantization and mapping method, Gallager mapping, etc. may be used to generate Gaussian-like signals \cite{Feng2015, Ma2004, Gadkari1999}.

In this paper, the transmissions are assumed to be Gaussian sources, but it does not mean that the proposed SA-GMP only works for Gaussian transmissions. The Gaussian source assumption is employed to prove that the SA-GMP converges to the optimal LMMSE detection. For non-Gaussian transmissions, the SA-GMP also works well but it is hard to prove the optimality of it. It is similar to the case of LMMSE detection which is optimal for the Gaussian source detection, but it also works well in many non-Gaussian cases. For example, in \cite{Lei20161}, it shows that the iterative LMMSE detector with a matched multiuser coding is sum-capacity achieving for the MIMO-NOMA system. Furthermore, our simulation results in Fig. \ref{f8} show that the proposed method in this paper also works well for the practical discrete MIMO-NOMA system.

\subsection{Iterative Receiver}
 We adopt a joint detection-decoding iterative receiver, which is widely used in the CDMA systems \cite{Wang1999} and the \emph{Inter-Symbol Interference} (ISI) channels \cite{Tuchler2011} for the overloaded massive MIMO-NOMA system. The messages ${\large{\textbf{\emph{e}}}_{\small{ESE}}}({\mathbf{x}}_{i})$, ${\tilde {\textbf{\emph{l}}}_{\small{ESE}}} ({\mathbf{x}}_{i})$, ${\tilde {\textbf{\emph{l}}}_{\small{DEC}}} ({\mathbf{c}}_{i})$, and ${\large{\textbf{\emph{e}}}_{\small{DEC}}}({\mathbf{c}}_{i})$, $i\in\mathcal{N}_u$, are defined as the input and output estimates of ${\mathbf{x}}_{i}$ at ESE and ${\mathbf{c}}_{i}$ the decoders. {As illustrated in Fig. \ref{f2}, at the BS, the received signals $\mathbf{Y}=[\mathbf{y}_1,\cdots,\mathbf{y}_N]$ and \emph{a-priori} message  $\{{\tilde {\textbf{\emph{l}}}_{\small{ESE}}} ({\mathbf{x}}_{i}),i\in \mathcal{N}_u\}$ are passed to a MIMO multi-user detector (MUD) called the elementary signal estimator (ESE) to estimate the MUD-extrinsic message ${\large{\textbf{\emph{e}}}_{\small{ESE}}}({\mathbf{x}}_{i})$ for each decoder $i$, which is then re-demodulated and re-deinterleaved (with $\Pi_i^{-1}$) into ${\tilde {\textbf{\emph{l}}}_{\small{DEC}}} ({\mathbf{c}}_{i})$, $i\in \mathcal{N}_u$. The corresponding single-user decoder calculates the decoder-extrinsic message ${\large{\textbf{\emph{e}}}_{\small{DEC}}}({\mathbf{c}}_{i})$ based on ${\tilde {\textbf{\emph{l}}}_{\small{DEC}}} ({\mathbf{c}}_{i})$. Then, this message is interleaved (by $\Pi_i$) and {re-modulated} to obtain new \emph{a-priori} message ${\tilde {\textbf{\emph{l}}}_{\small{ESE}}} ({\mathbf{x}}_{i})$ for the ESE. This process is repeated iteratively until the maximum number of iterations is achieved.} In fact, the messages ${\large{\textbf{\emph{e}}}_{\small{ESE}}}({\mathbf{x}}_{i})$, and ${\tilde {\textbf{\emph{l}}}_{\small{ESE}}} ({\mathbf{x}}_{i})$  can be replaced by the means and variances respectively if the messages are all Gaussian distributed.

 In the paper, we consider the low-complexity GMP as the ESE for the iterative receiver. Before discussing the GMP, we first present some results of the LMMSE estimator. These results will be used to support the convergence analysis and performance comparison for the GMP in the rest of this paper. %\vspace{0.2cm}
%\subsubsection{Decoding}
%The output message of the decoder is defined as%\vspace{-0.3cm}
%\begin{equation}\label{e16}
%{\large{\emph{e}}}_{\small{DEC}}({{x}}'_{i,t})=p(x'_{i,t}|{\tilde {\textbf{\emph{l}}}_{\small{DEC}}} ({\mathbf{x}}'_{ i,\sim t})),%\vspace{-0.3cm}
%\end{equation}
%where $i\in \mathcal{N}_u$, $t\in \mathcal{N}$, and ${\tilde {\textbf{\emph{l}}}_{\small{DEC}}} ({\mathbf{x}}'_{ i,\sim t})$ represents the vector obtained by deleting the $t$th entry of $[{\tilde{{\emph{l}}}_{\small{DEC}}}({{x}}'_{i,1}),\cdots, {\tilde{{\emph{l}}}_{\small{DEC}}}({{x}}'_{i,N})]$.

\subsection{LMMSE Estimator}
In the massive MIMO-NOMA, the complexity of the optimal MAP estimator is too high, and the LMMSE estimator is an alternative low-complexity ESE. For the massive MIMO-NOMA system, the LMMSE estimator is an optimal linear detector \cite{verdu1998} when the sources are Gaussian distributed, and is sum-capacity achieving with a matched multiuser coding \cite{Lei20161}. Let $\bar{\mathbf{x}}^l_t=[{\bar{x}_{1,t}^l},\cdots,\bar{x}_{N_u,t}^l]$ and $\mathbf{V}^l_{\bar{\mathbf{x}}_t}=\mathbf{V}^l_{\bar{\mathbf{x}}}= \mathrm{diag}(\bar{v}_1^l,\bar{v}_2^l,\cdots,\bar{v}_{N_u}^l)$ denote the expectation and variance of the prior message ${\tilde {\textbf{\emph{l}}}_{\small{ESE}}} ({\mathbf{x}}(t))$. The LMMSE detector \cite{tse2005} is
\begin{equation}\label{GMP2}
{{\hat {\mathbf{x}}}_t} = \mathbf{V} _{{{\hat {\mathbf{x}}}}}^{-1}\left[\mathbf{V}_{\bar{\mathbf{x}}}^{l^{-1}}\bar{\mathbf{x}}_t^l+ \sigma^{-2}_n\mathbf{H}^T\mathbf{y}_t  \right],%\vspace{-0.25cm}
\end{equation}
where $\mathbf{V} _{{{\hat {\mathbf{x}}}}} = (\sigma _{{{n}}}^{- 2}\mathbf{H}^T\mathbf{H}+\mathbf{V} _{{{ \bar{\mathbf{x}}}}}^{l^{-1}})^{-1}$. The extrinsic LMMSE estimation of $x_{i,t}$ is calculated by excluding the contribution of the \emph{a-priori} message $[\bar{x}_{i,t}^l,\bar{v}_{i}^l]$ according to the Gaussian message combining rule \cite{Loeliger2004,Loeliger2006} as follows.\vspace{-0.3cm}
\begin{equation}\label{e9}
\bar{v}_i^{e^{-1}} = \hat{v}_{i}^{-1}-\bar{v}_i^{l^{-1}}, \;\;\mathrm{and}\;\; \frac{\bar{x}_{i,t}^e}{\bar{v}_i^{e} }=\frac{\hat{x}_{i,t}}{\hat {v}_{i}}-\frac{\bar{x}^l_{i,t}}{\bar{v}_i^{l}},%\vspace{-0.4cm}
\end{equation}
where $i\in \mathcal{N}_u$, $t\in \mathcal{N}$, and $[\hat{x}_{i,t},\hat{v}_i]$ is the entry of $[{{\hat {\mathbf{x}}}_t},\mathbf{V} _{{{\hat {\mathbf{x}}}}}]$.

From (\ref{GMP2}), the MSE of LMMSE detector is calculated by\vspace{-0.3cm}
\begin{equation}\label{PA5}
MSE_{\small mmse}=\tfrac{1}{N_u}{\rm{tr}}(\mathbf{V}_{\hat {\mathbf{x}}}) = \frac{1}{N_u}{\rm{tr}}\{(\sigma _{{{n}}}^{ - 2}\mathbf{H}^T\mathbf{H}\!+\!\mathbf{V} _{{{ {\bar{\mathbf{x}}}}}}^{-1})^{-1}\}.%\vspace{-0.2cm}
\end{equation}

\subsubsection{MSE of LMMSE Estimator}
The following proposition is obtained from random matrix theory\footnote{We first consider  $\bf{y}'=\bf{A}\bf{x}+\bf{n'}$ where $\mathbf{A}\sim\mathcal{N}^{N_s\times N_u}(0,1/N_s)$ and ${\bf{n'}}\sim(0,\sigma_n^2/N_s)$. According to the results in \cite{verdu2004}, Eqn. (5) gives an asymptotic MSE estimation for the LMMSE detection of $\bf{y}'=\bf{A}\bf{x}+\bf{n'}$ (by treating the variance of noise as $\sigma_n^2/N_s$). Then, $\bf{y}=\bf{H}\bf{x}+\bf{n}$ can be obtained by scaling both sides of the equation with $\sqrt{N_s}$. Since the scaling (multiplying or dividing) operation does not change the MSE of LMMSE detection. Thus, Eqn. (5) also gives an asymptotic MSE estimation for the LMMSE detection of $\bf{y}=\bf{H}\bf{x}+\bf{n}$.
} \cite{verdu2004, Lei2016}.\vspace{0.25cm}

\textbf{\textit{Proposition 1:}} \emph{In the massive MIMO-NOMA, where $\beta= N_u/N_s$ is fixed, $N_u$ is large, and $\mathbf{V}_{\bar{\mathbf{x}}}^l=\bar{v}^lI_{N_u}$, the MSE (or \emph{a-posteriori} variance) of the LMMSE detector is given by%\vspace{-0.2cm}
\begin{eqnarray}\label{PA6}
\hat{v}_i \approx \hat{v}_{mmse}\!=\! \bar{v}^l \!\!-\! \frac{\sigma_n^2}{{4N_u}}\Big( \sqrt {snr^lN_s{{\left( {1 \!+\! \sqrt \beta } \right)}^2} \!\!+\! 1} \qquad\quad\nonumber\\
\qquad\qquad\quad -\sqrt {snr^lN_s{{\left( {1 \!\!-\! \sqrt \beta } \right)}^2} \!\!+\! 1}  \Big)^2\!\!,%\vspace{-0.35cm}
\end{eqnarray}
where $i\in {\mathcal{N}}_u$, and $snr^l={{\bar{v}^l} \mathord{\left/
 {\vphantom {{\bar{v}^l} {\sigma _n^2}}} \right.
 \kern-\nulldelimiterspace} {\sigma _n^2}}$ is the signal-to-noise ratio.}

\textbf{\textit{Remark 1}}: Although (\ref{PA6}) is an asymptotic result, it also comes close to the actual MSE of the MIMO-NOMA system with small values of $N_u$ and $N_s$ \cite{verdu1998}.
\subsubsection{Comparison Between Coded and Uncoded MIMO-NOMA}
In the uncoded MIMO-NOMA, the input variance is fixed during the iteration, i.e., $\bar{v}^l=\sigma_x^2$. In this case, for an overloaded system with $\beta>1$, the MSE of the LMMSE detector is $\frac{{N_u - N_s}}{N_u}\sigma_x^2$, which is independent of the Gaussian noise and has a poor MSE that is very close to $\sigma_x^2$. The reason is that when $N_u>N_s$, the inter-user interference limits the system performance. Hence, we introduce the the error-correcting code to mitigate the errors introduced by the inter-user interference in the overloaded massive MIMO-NOMA. This can be explained from the perspective of information theory. Although, under the power constraint $P_u=\sigma_x^2$, the sum capacity of the overloaded MU-MIMO system ($R_{sum}=\log \det(I_{N_s} + \frac{P_u}{\sigma_n^2}\mathbf{H}\mathbf{H}^T)$) increases with $N_u$, the average user rate ($R_{u}=\frac{1}{N_{u}}\log \det(I_{N_s} + \frac{P_u}{\sigma_n^2}\mathbf{H}\mathbf{H}^T)$) decreases to zero when $N_u\to \infty$ and $\beta= N_u/N_s>1$. Error-correcting codes are employed to decrease the user rate to meet the capacity requirement of the overloaded massive MIMO-NOMA system.

In the coded massive MIMO-NOMA, the variance $\bar{v}^l$ is decreasing with the increase of the iterations with the decoders. As a result, the system performance is improved through the joint iteration between the LMMSE detector and single-user decoders. We use a simple repetition code (error-correcting code) as an example, in which each user transmits every symbol $R$ times. This is equivalent to have $RN_s$ antennas at the BS. If $R>N_u/N_s$, the overloaded system is degenerated to an underloaded system ($N_u<RN_s$).

\subsubsection{Complexity of LMMSE Estimator} The complexity of the LMMSE estimator is $\mathcal{O}((N_u^3+N_sN_u^2)N_{ite})$, where $N_{ite}$ is the number of iterations. When $N_u$ and $N_s$ are large, the complexity of LMMSE estimator is too high to be practical. Hence, designing a low-complexity detector with little or no performance loss for the overloaded massive MIMO-NOMA is important. In this paper, we consider the low-complexity GMP for this purpose.}

\section{Gaussian Message Passing Detector and Convergence Analysis}
\begin{figure}[t]
  \centering
  \includegraphics[width=8.5cm]{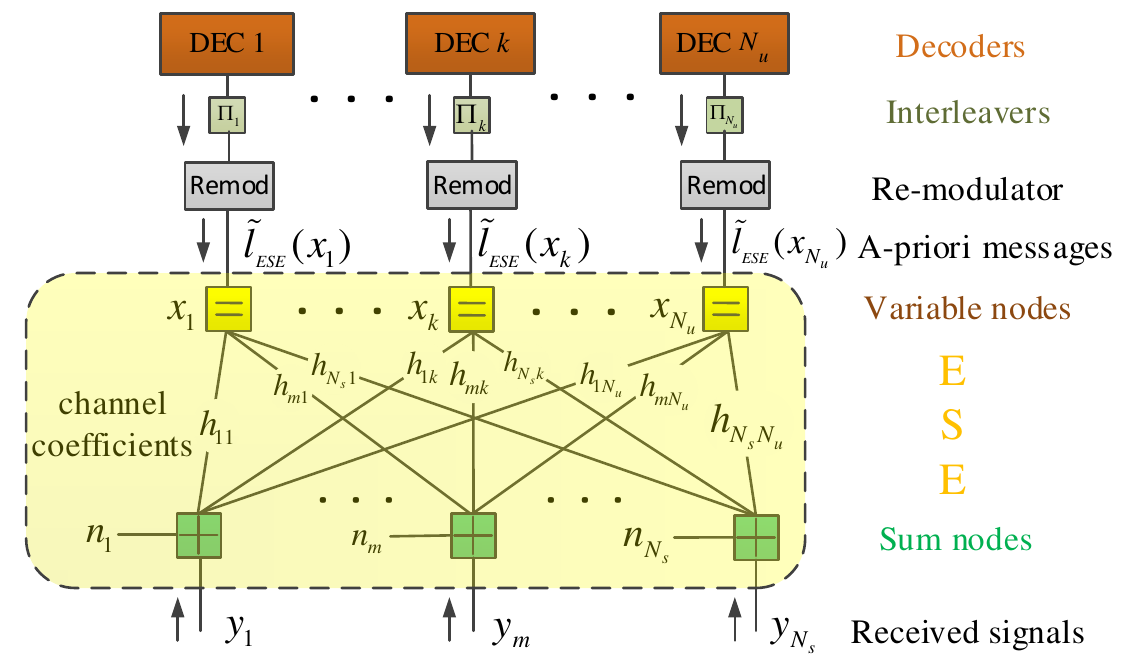}\\
  \caption{Factor graph of ESE (i.e. multi-user detector) for MIMO-NOMA. With the input $\mathbf{y}$ and $\{{\tilde {\textbf{\emph{l}}}_{\small{DEC}}} ({\mathbf{x}}_{i}),i\in \mathcal{N}_u\}$, GMP updates and passes Gaussian messages iteratively on this factor graph. Note that the channel coefficients $\{h_{mi}\}$ are used to multiply or normalize the messages passed between the sum nodes and variable nodes.}\label{fac_graph}
\end{figure}

Fig. \ref{fac_graph} shows a bipartite factor graph of the MIMO-NOMA system discussed in this paper. The GaBP \cite{Roy2001,Loeliger2006} and its asymptotic version AMP \cite{Donoho2009,Donoho2010a} are the state-of-art MPAs for MIMO detection. However, they have convergence difficulty under certain system loads. In this paper, we present the GMP and analyze its convergence for the overloaded massive MIMO-NOMA. GMP is similar to the GaBP in that Gaussian messages are passed on the edges of the Gaussian factor graph. However, while GaBP passes the extrinsic message on the whole graph, GMP updates the \emph{a-posteriori} messages at the sum nodes. In the rest of this paper, we replace the ESE in Fig. \ref{f2} with the GMP, and we drop the subscript $t$ for simplicity.

\subsection{Sum-Node Message Update of GMP}
Each SN can be seen as a multiple-access process
and its message is updated by
\begin{equation}\label{np1}
\left\{ \begin{array}{l}
x_{m \to k}^s(\tau) = {y_m} - \sum\limits_{i} {h_{mi}x_{i \to m}^u(\tau-1)\;,} \\
v_{m \to k}^s(\tau)= \sum\limits_{i} {h_{mi}^2v_{i \to m}^u(\tau-1) + \sigma _n^2\;},
\end{array} \right.%\vspace{-0.2cm}
\end{equation}
where $i,k \in \mathcal{N}_u ,m \in \mathcal{N}_s$, $\tau$ denotes the $\tau$th iteration, $y_m$ is the $m$th entry of $\textbf{{y}}$, $h_{mi}$ is the entry of $\mathbf{H}$ in $m$th row and $i$th column, and $\sigma^2_n$ denotes the variance of the Gaussian noise. In addition, $x_{k \to m}^u(\tau)$ and $v_{k \to m}^u(\tau)$ denote the mean and variance passing from the $k$th VN to $m$th SN respectively, and $x_{m \to k}^s(\tau)$ and $v_{m \to k}^s(\tau)$ denote the mean and variance passing from $m$th SN to $k$th VN respectively. The initial values of $v_{k\to m}^u(0)$ and $x_{k\to m}^u(0)$ are $+\infty$ and $0$, respectively. Different from GaBP, the message update at SN outputs a-posteriori messages to its connected edges.
\subsection{Variable-Node Message Update of GMP}
Each VN is a broadcast process and updated by%\vspace{-0.2cm}
\begin{equation}\label{var_mes}
\!\!\!\!\left\{ \!\!\!\!{\begin{array}{*{20}{l}}
{v_{k \to m}^u(\tau) \!=\! {{( {\sum\limits_{i\ne m} {h_{ik}^2v_{i \to k}^{s{\;^{ - 1}}}(\tau) + \bar{v}_{{k}}^{ l^{-1}}\;} } )}^{ - 1}},}\\
{x_{k \to m}^u(\!\tau\!) \!=\! v_{k \to m}^u(\!\tau\!)(\sum\limits_{i \ne m} \!\!{{h_{ik}}v_{i \to k}^{s{\;^{ - 1}}}(\!\tau\!)x_{i \to k}^s(\!\tau\!)} \!+\! \bar{v}_{{k}}^{ l^{-1}}\bar{x}_k^{l}) }.
\end{array}} \right.
\end{equation}
where $k \in \mathcal{N}_u, i,m \in \mathcal{N}_s$, and $\bar{v}_{{k}}^{ l}$ and $\bar{x}_k^{l}$ denote the estimated variance and mean of $x_k$ from the decoder $k$ respectively. {The message update at SN outputs extrinsic messages, which is the same as that in GaBP.}

\subsection{Extrinsic message output and Decision of GMP}
The GMP outputs the extrinsic messages:
\begin{equation}\label{np2}
\left\{ \begin{array}{l}
\bar{v}^e_{k} = {(\, {\sum\limits_{m } {h_{mk}^2v_{m \to k}^{s{\;^{ - 1}}}(\tau)\,} } )^{ - 1}},\\
\bar{x}_{k}^e = \bar{v}^e_{k}\sum\limits_{m } {h_{mk}v_{m \to k}^{s{\;^{ - 1}}}(\tau)x_{m \to k}^s(\tau)\;\;},
\end{array} \right.
\end{equation}
where $k \in \mathcal{N}_u, m \in \mathcal{N}_s$. When the MSE of GMP meets the requirement or the number of iterations reaches the limit, we output the \emph{a-posteriori} estimation $\hat{x}_{k}$ and its MSE $\hat{v}_{k}$:
\begin{equation}\label{np3}
\left\{ \begin{array}{l}
\hat{v}_{k} = {(\, {\sum\limits_{m } {h_{mk}^2v_{m \to k}^{s{\;^{ - 1}}}(\tau)\,} }+\bar{v}_{{k}}^{ l^{-1}} )^{ - 1}},\\
\hat{x}_{k} = \hat{v}_{k}(\sum\limits_{m } {h_{mk}v_{m \to k}^{s{\;^{ - 1}}}(\tau)x_{m \to k}^s(\tau) + \bar{v}_{{k}}^{ l^{-1}}\bar{x}_k^{l}}).
\end{array} \right.\quad
\end{equation}

\textbf{\emph{Remark 2}}: In the sum-node message update, GMP passes the \emph{a-posteriori} information, not the extrinsic information (GaBP), from the SNs to the VNs.We show in Section III.F that the variances of the \emph{a-posteriori} message update and the extrinsic message update converge exactly to the same value, but their means behave differently. It is hard to analyse the mean convergence of GaBP because of its complicated structure. However, the convergence conditions of the mean of GMP can be derived. The reason is that, when using \emph{a-posteriori} message update at the SN, the GMP converges to a classical iterative algorithm, whose convergence depends on {a spectral radius}. Based on these points, this paper proposes an optimized SA-GMP to minimize the spectral radius, thus obtaining better MSEs and convergence properties (e.g. the fixed point, convergence condition and speed) than GMP and AMP/GaBP (see Figs. \ref{f5}-\ref{fig_AMP}).

Another point to note is that, in the underloaded GMP and underloaded SA-GMP schemes considered in \cite{Lei2016}, the VNs output \emph{a-posteriori} messages, and the SNs output extrinsic messages. However, the \emph{a-posteriori} message update at VNs in \cite{Lei2016} will lead to higher performance loss when the MIMO NOMA system is overloaded (i.e. the number of users is larger than that of antennas), because the \emph{a-priori} information (on each edge) at the variable node account for a larger proportion in the \emph{a-posteriori} information than that at the sum node, which in turn aggravate the correlation problem in GMP. Therefore, in the proposed SA-GMP, we let the VNs output the extrinsic messages and the SNs output the \emph{a-posteriori} messages to mitigate the correlation problem.

Here we provide an intuition for the \emph{a-posteriori} and extrinsic message update. Generally, extrinsic message processing is used to cut off the loops in a loopy factor graph. However, it is shown that extrinsic message update is not necessary for all the processors in a loop. For example, in this paper, for a {bipartite} loopy factor graph, it only needs extrinsic update at the VNs and \emph{a-posteriori} update at the SNs. As a result, the loops in the {bipartite} factor graph are cut off by the partial extrinsic update at the VNs. In detail, in this paper, the variance of GMP converges to the MSE of LMMSE, because $\sum_{i\ne m} {h_{ik}^2}$ (extrinsic update at the VNs) and $\sum_{j} {h_{ij}^2}$ (\emph{a-posteriori} update at the SNs) are independent with each other\footnote{Similarly, in \cite{Lei2016}, on a {bipartite} loopy factor graph, {underloaded GMP} only updates extrinsic messages at the SNs, and updates \emph{a-posteriori} messages at the VNs. In addition,  $\sum_{i} {h_{ik}^2}$ (\emph{a-posteriori} update at the VNs) and $\sum_{j\ne k} {h_{ij}^2}$ (extrinsic update at the SNs) are independent with each other.}, which are the same as the case that extrinsic update is used at both the VNs and the SNs, i.e., $\sum_{i\ne m} {h_{ik}^2}$ and $\sum_{j\ne k} {h_{ij}^2}$ are independent with each other. Apart from that, in the mean update of GMP, the partial extrinsic update introduces the term $\mathbf{D}_{\mathbf{H}{\mathbf{H}^T}}$ in (\ref{ne_matr_f}), which greatly reduces {the spectral radius} so that the GMP can converge in many cases. These are why the partial extrinsic update in each loop works for the GMP. However, if we use \emph{a-posteriori} update at both the VNs and the SNs, $\sum_{i} {h_{ik}^2}$ and $\sum_{j} {h_{ij}^2}$ in the variance update will not be independent with each other as they have a common term ${h_{ik}^2}$. Hence, the variance of GMP will not converge to the MSE of LMMSE. In this case, the $\mathbf{D}_{\mathbf{H}{\mathbf{H}^T}}$ in the mean update also disappears, which greatly increases {the spectral radius} and makes the GMP diverge in most cases. These are the reasons why a full \emph{a-posteriori} update fails to work for the loopy factor graphs.

\subsection{GMP in Matrix Form}
{Let $\mathbf{A}_{M\times N}.*\mathbf{B}_{M\times N}=\left[ a_{ij}b_{ij}\right]_{M\times N}$, $diag^{-1}\{\mathbf{A}_{N\times N}\}=\left[a_{11},a_{22},\cdots,a_{NN}\right]^T$, $\mathbf{1}_{M\times N}=\left[1\right]_{M\times N}$, and $\mathbf{A}^{(k)}_{M\times N}=\left[a^k_{ij}\right]_{M\times N}$. Assume $\mathbf{X}_{su}(\tau)=\left[x_{m\to i}^s(\tau)\right]_{N_s\times N_u}$, $\mathbf{V}_{su}(\tau)=\left[v_{m\to i}^s(\tau)\right]_{N_s\times N_u}$, $\mathbf{X}_{us}(\tau)=\left[x_{i\to m}^u(\tau)\right]_{N_u\times N_s}$, and $\mathbf{V}_{us}(\tau)=\left[v_{i\to m}^u(\tau)\right]_{N_u\times N_s}$. Algorithm 1 shows the detailed process of the GMP. Message update (\ref{np1}) is rewritten to step 5, where $\widetilde{\mathbf{X}}_{us}(\tau)=\mathbf{X}_{us}(\tau).*\mathbf{H}^T$ and $\widetilde{\mathbf{V}}_{us}(\tau)=\mathbf{V}_{us}(\tau).*{\mathbf{H}^{(2)}}^{T}$.
Let $\bar{\mathbf{v}}^l_{\mathbf{x}}=[\bar{v}_1^l,\ldots, \bar{v}_{N_u}^l]^T$, $\bar{\mathbf{x}}^l=[\bar{x}_1^l,\ldots,\bar{x}_{N_u}^l]^T$, $\mathbf{W}_{su}(\tau)=\mathbf{V}_{su}^{(-1)}(\tau)$, $\mathbf{W}_{us}(\tau)={\mathbf{V}_{us}^{(-1)}(\tau)}$, and $\mathbf{G}_{us}(\tau)=\mathbf{W}_{us}(\tau).*\mathbf{X}_{us}(\tau)$. Message update (\ref{var_mes}) is rewritten to step 7, where $\widetilde{\mathbf{W}}_{su}(\tau)=\mathbf{H}^{(2)}.*\mathbf{W}_{su}(\tau)$ and $\widetilde{\mathbf{G}}_{su}(\tau)=\mathbf{H}.*\mathbf{G}_{su}(\tau) = \mathbf{H}.*\mathbf{W}_{su}(\tau).*\mathbf{X}_{su}(\tau)$.
 We have $\mathbf{V}_{us}(\tau)$ from $\mathbf{V}_{us}(\tau)=\mathbf{W}_{us}^{(-1)}(\tau)$, and $\mathbf{X}_{us}(\tau)$ from $\mathbf{X}_{us}(\tau)=\mathbf{V}_{us}(\tau).*\mathbf{G}_{us}(\tau)$. In step 10, we let $\bar{{\mathbf{x}}}^e=[\bar{x}_1^e,\ldots,\bar{x}_{N_u}^e]$ and $\bar{\mathbf{v}}^e=[\bar{v}_1^e,\ldots,\bar{v}_{N_u}^e]$.}
\vspace{-0.2cm}
\begin{algorithm}[t]
\caption{GMP Algorithm}
\begin{algorithmic}[1]
\State {\small{\textbf{Input:} {{$\bar{\mathbf{x}}^l$, $\bar{\mathbf{v}}^l_{\mathbf{x}}$, $\sigma^2_n$, $\epsilon>0$, $N_{ite}^{ese}$}}, $\mathbf{H}$ and calculate {{$\mathbf{H}^{(2)}$}}.
\State \textbf{Initialization:} {{$\tau\!=\!0$, $\mathbf{X}_{us}(0)\!=\!\bf{0}$}}, {{$\mathbf{V}_{us}(0)=+\boldsymbol{\infty}$.}}
\State \textbf{Do}
\State \quad\;\;$\tau\!=\!\tau\!+\!1$,
\State \quad\;\; $\widetilde{\mathbf{X}}_{us}(\tau)\!=\!\mathbf{X}_{us}(\tau).*\!\mathbf{H}^T$ and $\widetilde{\mathbf{V}}_{us}(\tau)\!=\!\mathbf{V}_{us}(\tau).\!*{\mathbf{H}^{(2)}}^{T}$,
\State  \vspace{-0.3cm}\[ \quad\begin{array}{l}
\left[\!\!\! \begin{array}{l}
{{\rm{\mathbf{X}}}_{su}}(\tau)\\
{\mathbf{V}_{su}}(\tau)
\end{array} \!\!\!\right]\!\! =\!\! \left[ \!\!\! \begin{array}{l}
\quad\;\;\textbf{{y}} - dia{g^{ - 1}}\{ \mathbf{1}_{M\times K}\cdot{\widetilde{\mathbf{X}}_{us}}(\tau- 1)\} \\
\sigma_n^2{\mathbf{1}}_{M\!\times\!1}\!+\!dia{g^{ - 1}}\{ \mathbf{1}_{M\!\times\! K}\!\cdot \!{\widetilde{\mathbf{V}}_{us}}(\tau \!-\! 1)\}
\end{array} \!\!\!\right]\!\! \cdot\!\! {\mathbf{1}_{1 \!\times\! K}},
\end{array}\]}
\State\vspace{-0.5cm} \[ \widetilde{\mathbf{W}}_{su}(\tau)\!=\!\mathbf{H}^{(2)}.*\mathbf{V}^{(-1)}_{su}(\tau), \qquad\qquad\qquad\qquad\qquad\] \[ \vspace{-0.2cm} \widetilde{\mathbf{G}}_{su}(\tau)\!=\!\mathbf{H}.*\mathbf{V}^{(-1)}_{su}(\tau).*\mathbf{X}_{su}(\tau), \qquad\qquad\qquad\quad\]
\State \vspace{-0.3cm}\[ \begin{array}{l}
\left[ \!\!\!\!\begin{array}{l}
{\mathbf{W}_{us}}(\tau)\\
{\mathbf{G}_{us}}(\tau)
\end{array} \!\!\!\!\right] \!\!=\!\!\! \left[ \!\!\!\begin{array}{l}
\bar{\mathbf{v}}^{l^{(-1)}}_{\mathbf{x}} \!\!+\! dia{g^{ - 1}}\left\{ {{\mathbf{1}} \!\cdot \!{{\widetilde{\mathbf{W}}_{su}}(\tau)}} \right\}\mathop {}\limits_{\mathop {}\limits_{} }\\
 \bar{\mathbf{v}}^{l^{(-1)}}_{\mathbf{x}}\!\!\!\!.*\!\bar{\mathbf{x}}^l \!\!+\! dia{g^{ - 1}}\!\!\left\{ {{\mathbf{1}} \!\cdot \! \widetilde{\mathbf{G}}_{su}(\tau)} \right\}
\end{array}\!\!\! \!\!\right] \!\cdot \! {\mathbf{1}}\!-\!\!\left[ \!\!\!\begin{array}{l}
\widetilde{\mathbf{{W}}}_{su}^T(\tau)\\
\widetilde{\mathbf{G}}_{su}^T(\tau)
\end{array} \!\!\!\right],
\end{array}\qquad\qquad\]\vspace{-0.2cm}
\State \quad\; $\mathbf{V}_{us}(\tau)=\mathbf{W}_{us}^{(-1)}(\tau)$ and  $\mathbf{X}_{us}(\tau)=\mathbf{V}_{us}(\tau).*\mathbf{G}_{us}(\tau)$.\vspace{0.2cm}
\State \textbf{While} \;{\small{$\left( \:|\mathbf{E}_{us}(\tau)-\mathbf{E}_{us}{(\tau-1)}|>\epsilon \;{\textbf{or}}\; \tau\leq N_{ite}^{ese} \;\right)$}}
\State \vspace{-0.3cm} {{ \[ \vspace{-0.1cm} \begin{array}{l}
\bar{\mathbf{v}}^e = {\left( {dia{g^{ - 1}}\left\{ {{\mathbf{1}_{K \times M}} \cdot {\widetilde{\mathbf{W}}_{su}}(\tau) } \right\}} \right)^{( - 1)}}\!\!\!\!,\\
\bar{{\mathbf{x}}}^e = \bar{{\mathbf{v}}}^e.*dia{g^{ - 1}}\left\{ {\mathbf{1}_{K \times M}} \cdot \widetilde{\mathbf{G}}_{su}(\tau) \right\}.
\end{array} \qquad\qquad\qquad\qquad\]}}
\State \textbf{Output:}  $\bar{{\mathbf{x}}}^e$ and $\bar{\mathbf{v}}^e $ }.
\end{algorithmic}
\end{algorithm}
\subsection{Complexity of GMP}
As the variance updates in (\ref{np1})$-$(\ref{np3}) are independent of the received $\textbf{{y}}$, they can be pre-computed before the iterative detection. Therefore, in each iteration, the GMP needs about $4N_uN_s$ multiplications and $4N_uN_s$ additions. Therefore, the complexity of GMP is as low as $\mathcal{O}(N_uN_sN_{ite}^{ese}N_{ite}^{out})$, where $N_{ite}^{ese}$ is the number of inner iterations at the GMP and $N_{ite}^{out}$ is the number of outer iterations between the ESE and decoders.

\subsection{Variance Convergence of GMP}%\vspace{-0.2cm}
{The variance convergence of GMP is shown by the following proposition, which is derived with the SE technique in \cite{Donoho2010a,Donoho2009,Bayati2011}.

\textbf{\textit{Proposition 2:}} \emph{In the massive MIMO-NOMA, where $\beta= N_u/N_s$ is fixed, $N_u$ is large, and $\mathbf{V}_{\bar{\mathbf{x}}}^l=\bar{v}^lI_{N_u}$, the \emph{a-posteriori} variances of GMP converge to
\begin{eqnarray}\label{p6}
\!\!\!\!\!\!\!\!\!\!\!\!\!\!\!&&\hat{v}_i\approx \!v_{i\rightarrow m}^u(\infty) = \hat{v}\!= \! MSE^{GMP}\nonumber\\
\!\!\!\!\!\!\!\!\!\!\!\!\!\!\!&&=\!\!\frac{{\sqrt {{{\!(snr^{-1} \!\!+ \!\!N_s \!\!-\!N_u)}^2} \!\! + \!4N_u snr^{-1}}  \!- \!(snr^{-1}\!\!+\!N_s \!\!-\! N_u)}}{{2N_u \bar{v}^{l^{-1}}}}\quad
\end{eqnarray}
where $i\in \mathcal{N}_u\,$, $m\in \mathcal{N}_s$, and $snr={\bar{v}^l \mathord{\left/
 {\vphantom {{\sigma _x^2} {\sigma _n^2}}} \right.
 \kern-\nulldelimiterspace} {\sigma _n^2}}$ is the signal-to-noise ratio.}

\begin{IEEEproof}
See APPENDIX A.
\end{IEEEproof}
}

It should be noted that as the user decoders are included at the receiver, the $snr$ can be very small and thus cannot be neglected in (\ref{p6}). From (\ref{PA6}) and (\ref{p6}), it is easy to verify that $\hat{v}=\hat{v}_{mmes}$, which means that the output extrinsic variances ($\bar{v}^e_i, i\in \mathcal{N}_u$) of the GMP and LMMSE detector are the same. Therefore, we obtain the following theorem.

\textbf{\textit{Theorem 1:}} \emph{In the massive MIMO-NOMA, where $\beta= N_u/N_s$ is fixed, $N_u$ is large, and $\mathbf{V}_{\bar{\mathbf{x}}}^l=\bar{v}^lI_{N_u}$, the variances of GMP converge to that of the LMMSE detector.}

From (\ref{np1}), $\{\mathbf{V}_{su}(\tau)\}$ also converges to a certain value, i.e., $v^s_{m\to k} \to v^s$ and
\begin{equation}\label{p8}
v^s \approx N_u{\hat v} + \sigma _n^2.
\end{equation}
Let $\gamma={{{\hat v}} \mathord{\left/
 {\vphantom {{{\hat v}} {{v^s}}}} \right.
 \kern-\nulldelimiterspace} {{v^s}}}$, from (\ref{p6}) and (\ref{p8}), we get
 {\begin{equation}\label{p7}
 \!\!\!\gamma  = \frac{1}{{{{N_u}} + {{\sigma _n^2} \mathord{\left/
 {\vphantom {{\sigma _n^2} {{{\hat v }}}}} \right.
 \kern-\nulldelimiterspace} {{{\hat v }}}}}} .
 \end{equation}}

\textbf{\emph{Remark 3}}:  In the original GMP, $\sum_{j}$ (or $\sum_{i}$) in the sum-node message update is replaced by $\sum_{j\neq k}$ (or $\sum_{i\neq k}$), and $N_u$ is thus replaced by $N_u-1$. However, when $N_u$ is large, $N_u-1\approx N_u$. Therefore, it is easy to find that the original GMP has the same results on the variance convergence. In addition, the above analysis provides an alternative way to estimate the MSE of the LMMSE detector.

\subsection{Mean Convergence of GMP}
{Previous work \cite{Lei2016} show that in underloaded massive MIMO-NOMA, the mean of {underloaded-GMP} converges to the LMMSE multi-user detector under a sufficient condition, and an {underloaded SA-GMP} whose mean and variance always converge to those of LMMSE multi-user detector with a faster convergence speed is proposed. However, for the overloaded case, the {underloaded GMP and underloaded SA-GMP} in \cite{Lei2016} have poor performance when overloaded, and the mean convergence analysis is different due to intractable interference between the large number of users.}

The following theorem gives two sufficient conditions for the mean convergence of the overloaded GMP. \vspace{0.25cm}
\subsubsection{Classical Iterative Algorithm}
We first introduce the classical iterative algorithm and its convergence proposition, which will be used for the mean convergence analysis of the GMP. The iterative algorithm \cite{Axelsson1994} is\vspace{-0.2cm}
\begin{equation}\label{a1}
{\textbf{\emph{x}}}(\tau) = \mathbf{B}\mathbf{x}(\tau - 1) + {\textbf{\emph{c}}},\vspace{-0.2cm}
\end{equation}
where neither the iteration matrix $\mathbf{B}$ nor the vector ${\textbf{\emph{c}}}$ depends upon the iteration number $\tau$.\vspace{0.25cm}

\textbf{\textit{Proposition 3 }}\cite{Bertsekas1989,Henrici1964}: \emph{Assuming that the matrix $\mathbf{I}-\mathbf{B}$ is invertible, the iteration (\ref{a1}) converges to the exact solution $\mathbf{x}^*=(\mathbf{I}-\mathbf{B})^{-1}\mathbf{c}$ for any initial guess $\mathbf{x}\left(0\right)$ if $\mathbf{I}-\mathbf{B}$ is strictly (or irreducibly) diagonally dominant or $\rho \left( \mathbf{B }\right) < 1$, where $\rho(\mathbf{B})$ is the spectral radius of $\mathbf{B}$.}\vspace{0.25cm}

\subsubsection{Mean Convergence of GMP} The following Lemma shows that the mean of GMP converges to the classical iterative algorithm.

\textbf{\textit{Lemma 1:}} \emph{In the overloaded massive MIMO-NOMA with $\mathbf{V}_{\bar{\mathbf{x}}}^l\!=\!\bar{v}^lI_{N_u}$, the sum-node messages of GMP satisfy $x_{ m\to k}^s(\!\tau\!)\!\!=\!\!x^s_{m}(\tau)$, $\!\forall m \!\!\in \!\!\mathcal{N}_s$, and converge to the following iterative algorithm.
\begin{equation}\label{p11}
\mathbf{x}^s(\tau) = \textbf{{y}} -  \gamma (\mathbf{H}\mathbf{H}^T-{\mathbf{D}_{{\mathbf{H}}\mathbf{H}^T}}) \mathbf{x}^s(\tau-1)-\alpha \mathbf{H}\bar{\mathbf{x}}^l,%\vspace{-0.2cm}
\end{equation}
where  $\mathbf{x}^s(\tau)={\left[ {{{x}_1^s}(\tau)\;{{x}_2^s}(\tau)\; \cdots \;{{x}_{N_s}^s}(\tau)} \right]^T}$.}
\begin{IEEEproof}
See APPENDIX B.
\end{IEEEproof}

Based on Lemma 1 and Proposition 3, we can have the following theorem.\vspace{0.25cm}

\textbf{\textit{Theorem 2:}} \emph{In the overloaded massive MIMO-NOMA, where $\beta= N_u/N_s$ is fixed, $N_u$ is large, and $\mathbf{V}_{\bar{\mathbf{x}}}^l=\bar{v}^lI_{N_u}$, the GMP converges to\vspace{-0.1cm}
\begin{equation}\label{f_gm}
\hat{\mathbf{x}}=\left( \theta \mathbf{H}^T\mathbf{H}+ \mathbf{I}_{N_u} \right)^{ - 1}\left( \theta \mathbf{H}^T\mathbf{y}+ \alpha\bar{\mathbf{x}}^l\right),\vspace{-0.1cm}
\end{equation}
where $\theta=\hat v/ \sigma^2_n$ and $\alpha=\hat{v}/\bar{v}^l$, if any of the following conditions holds.}

\emph{1. The matrix $\mathbf{I}_{N_s} + \gamma\left({\mathbf{H}}\mathbf{H}^T-\mathbf{D}_{{\mathbf{H}}\mathbf{H}^T}\right)$ is strictly or irreducibly diagonally dominant,}

\emph{2. $\rho \left( {\gamma ({\mathbf{H}}\mathbf{H}^T - {\mathbf{D}_{{\mathbf{H}}\mathbf{H}^T}})} \right) < 1$,}

\emph{\!\!\!\!\!\!where  $\gamma=\hat{v}/v^s$.}
\begin{IEEEproof}
See APPENDIX C.
\end{IEEEproof}

Comparing (\ref{GMP2}) with (\ref{f_gm}), we can see that the GMP converges to the LMMSE if $\hat v=\bar{v}^l$. However, from (\ref{np3}) or (\ref{p6}), we can see that $\hat v< \bar{v}^l$. Therefore, different from the underloaded case that the GMP converges to the LMMSE detection if it is convergent, the GMP in the overloaded massive MIMO-NOMA does not converge to the LMMSE detection even if it is convergent.

\subsubsection{ Spectral Radius and Convergence Point}
As $N_u$ and $N_s$ are large, $\beta>1$ and $\gamma=\mathcal{O}(\frac{1}{N_u})$ (see (\ref{p7})), from \emph{Random Matrix Theory}, we have
\begin{eqnarray}\label{radius1}
&&\rho_{_{GMP}} = \rho(\gamma(\mathbf{H}\mathbf{H}^T-\mathbf{D}_{\mathbf{H}\mathbf{H}^T}))\nonumber\\
&&\to \gamma N_u(\beta^{-1}+2\sqrt{\beta^{-1}})=\gamma(N_s+2\sqrt{N_sN_u}).\qquad\qquad
\end{eqnarray}
Then, we have the following corollary based on Theorem 2.

{\textbf{\emph{Corollary 1:}} \emph{In the overloaded massive MIMO-NOMA, where $\beta= N_u/N_s$ is fixed, $N_u$ is large, and $\mathbf{V}_{\bar{\mathbf{x}}}^l=\bar{v}^lI_{N_u}$, {the spectral radius} is given by
\begin{equation}
\rho_{_{GMP}} =\gamma(N_s+2\sqrt{N_sN_u}).
\end{equation}
If $\beta >(\sqrt{2}-1)^{-2}$, the GMP converges to
\begin{equation}
\hat{\mathbf{x}}=\left( \theta \mathbf{H}^T\mathbf{H}+ I_{N_u} \right)^{ - 1}\left( \theta \mathbf{H}^T\textbf{{y}}+ \alpha\bar{\mathbf{x}}^l\right),
\end{equation}
where $\gamma=(N_u+\sigma_n^2/\hat{v})^{-1}$, and $\hat{v}$ is given in (\ref{p42}).}}\vspace{0.25cm}

According to Theorem 2 and Corollary 1, it is easy to find that the convergence condition of the GMP (e.g., $\beta >(\sqrt{2}-1)^{-2}$ or $\rho \left( {\gamma ({\mathbf{H}}^T\mathbf{H} - {\mathbf{D}_{{\mathbf{H}}^T\mathbf{H}}})} \right) < 1$ ) in the overloaded massive MIMO-NOMA is also different from the underloaded case, which requires that $\beta <(\sqrt{2}-1)^{2}$ or $\rho \left( {\gamma ({\mathbf{H}}\mathbf{H}^T - {\mathbf{D}_{{\mathbf{H}}\mathbf{H}^T}})} \right) < 1$.

Let $\bigtriangleup \mathbf{x}(\tau)=\mathbf{x}^*-\mathbf{x}^s(\tau)$ be the mean deviation vector. From (\ref{p11}), we have
\begin{equation}\label{conv_speed}
\Delta \mathbf{x}(\tau) = \gamma \left( {{\mathbf{H}}\mathbf{H}^T - {\mathbf{D}_{{\mathbf{{H}}}\mathbf{H}^T}}} \right)\Delta \mathbf{x}(\tau - 1).
\end{equation}
Therefore, the means converge to the fixed point at an exponential rate of $\rho _{{\gamma ({\mathbf{H}}\mathbf{H}^T - {\mathbf{D}_{{\mathbf{H}}\mathbf{H}^T}})} }^{\tau}$, i.e., the smaller spectral radius is, the faster convergence speed it has.

\subsubsection{Comparison with LMMSE} Comparing the GMP (\ref{f_gm}) with LMMSE detection (\ref{GMP2}), we obtain the following corollary.

\textbf{\emph{Corollary 2:}} \emph{In the overloaded massive MIMO-NOMA, even if the GMP converges, it converges to a value with a worse MSE than that of the LMMSE detection.}
\begin{IEEEproof}
See APPENDIX D.
\end{IEEEproof}

Corollary 2 denotes that GMP is worse than the LMMSE detection even if it converges.
\begin{figure*}[t]
  \centering
  \includegraphics[width=13cm]{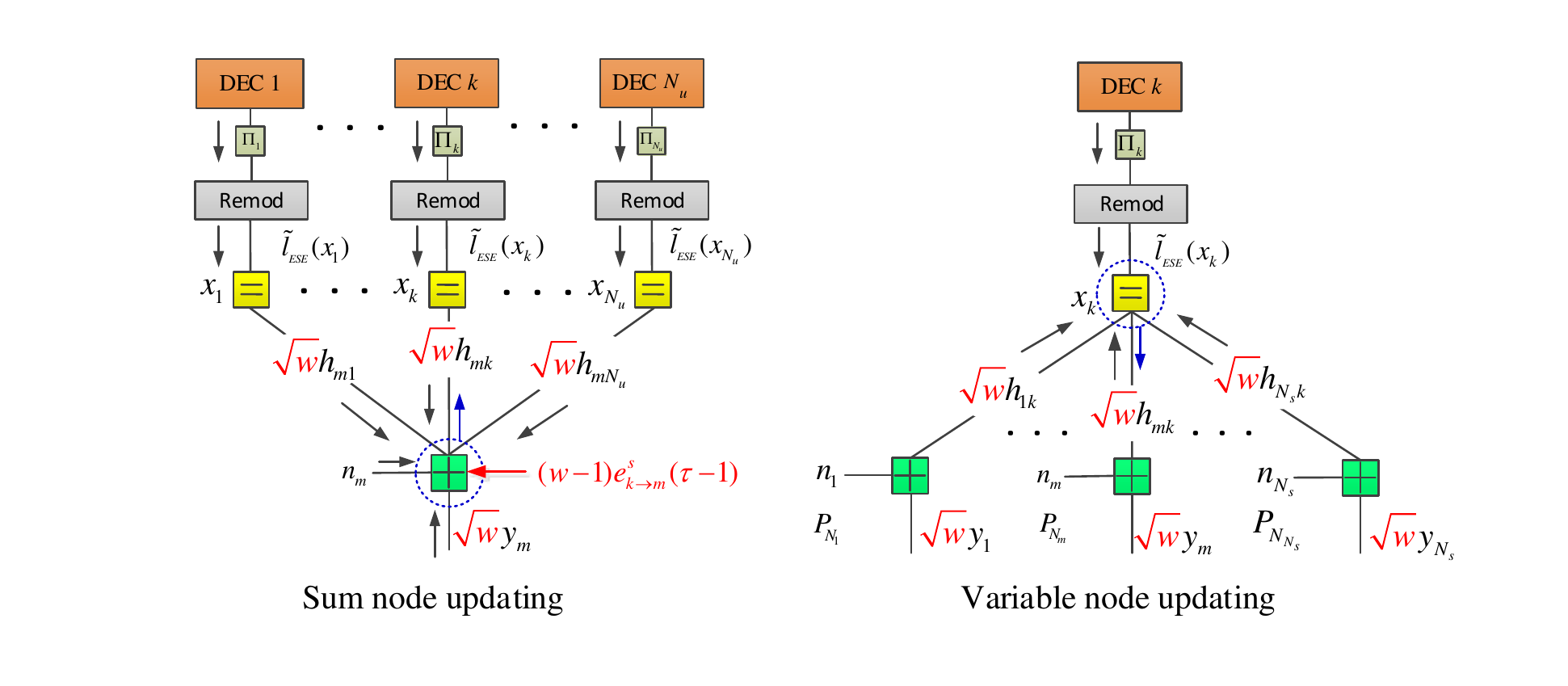}\\
  \caption{Mean update modification at the SNs and VNs of the proposed SA-GMP: the received $\mathbf{y}$ and channel matrix $\mathbf{H}$ are scaled with a relaxation parameter $w$, and a new term $(w-1)e^{s}_{k\to m}(\tau-1)$ is added in the SN update. However, the variance update remains unchanged from GMP. The scaling and addition terms introduced by SA-GMP (not present in the traditional GMP) are printed in red color font. The blue arrows denote the output messages generated by the sum node and variable node, given by Expressions (\ref{FC2}) and (\ref{FC1}) respectively.} \label{sa_GMP}
\end{figure*}
\section{A New Fast-Convergence SA-GMP that Approaches LMMSE Performance}
As shown in Section III, the GMP does not converge to the optimal LMMSE detection and has a low convergence speed. The main reason is that {the spectral radius} does not achieve the minimum value. Therefore, we propose a new fast-convergence scale-and-add GMP (SA-GMP). As shown in Fig. \ref{sa_GMP}, the SA-GMP is obtained by modifying the mean updates of GMP with linear operators, those are: {1)} scaling the received $\mathbf{y}$ and the channel matrix $\mathbf{H}$, i.e., ${\mathbf{H}'}=\sqrt{w}\mathbf{H}$ and $\textbf{{y}}'=\sqrt{w}\textbf{{y}}$, where $h'_{mk}=\sqrt{w}h_{mk}$ is an element of matrix $\mathbf{H}'$; \emph{2)} adding a new term $\!(\!w\!-\!1\!)x_{m \to k}^s\!(\!\tau\!-\!1\!)$ on the mean message update at each SN. However, we keep the variance output at the VNs the same as that of the GMP, because it always converges to the variance of the optimal LMMSE detection. By doing so, we can optimize the relaxation parameter $w$ to minimize {the spectral radius}.%\vspace{-0.3cm}

\subsection{Sum-Node Message Update of SA-GMP}
As shown in the left subfigure of Fig. \ref{sa_GMP}, the sum-node message update (\ref{np1}) is changed to
\begin{equation}\label{FC2}
\!\!\!\!\!\!\!\left\{\!\!\!\! \begin{array}{l}
x_{m \to k}^s(\!\tau\!) \!=\! {y'_m} \!\!-\!\! \sum\limits_{i} {h'_{mi}x_{i \to m}^u(\!\tau\!\!-\!\!1\!)}\!-\!\underbrace{(w\!\!-\!\!1)x_{m \to k}^s(\!\tau\!\!-\!\!1\!)}_{\mathrm{new \; term}}, \\
v_{m \to k}^s(\tau)= \sum\limits_{i} {h_{mi}^2v_{i \to m}^u(\tau-1) + \sigma _n^2\;},
\end{array} \right.%\vspace{-0.2cm}
\end{equation}
for $i, k \in \mathcal{N}_u,m \in \mathcal{N}_s$.%\vspace{-0.5cm}

\subsection{Variable-Node Message Update of SA-GMP}
As shown in the right subfigure of Fig. \ref{sa_GMP}, the variable-node message update (\ref{var_mes}) is changed to
\begin{equation}\label{FC1}
\!\!\!\!\!\!\left\{ {\begin{array}{*{20}{l}}
{v_{k \to m}^u(\tau) = {{( {\sum\limits_{i\ne m} {h_{ik}^2v_{i \to k}^{s{\;^{ - 1}}}(\tau) + \bar{v}_{{k}}^{ l^{-1}}\;} } )}^{ - 1}},}\\
{x_{k \to m}^u(\tau) = \bar{v}_{k}^l(\sum\limits_{i \ne m} {{h'_{ik}}\bar{v}_{i }^{s{\;^{ - 1}}}x_{i \to k}^s(\tau)} + \bar{v}_{{k}}^{ l^{-1}}\bar{x}_k^{l}) },
\end{array}} \right.\quad%\vspace{-0.2cm}
\end{equation}
where $\bar{v}_m^s= \sum\limits_{k} {h_{mk}^2\bar{v}_{k}^l + \sigma _n^2}$,
and $k \in \mathcal{N}_u, i,m \in \mathcal{N}_s$.

\subsection{Extrinsic message output and Decision of SA-GMP}
The extrinsic estimation $\bar{x}_{k}^e$ and its variance $\bar{v}^e_{k}$ of SA-GMP are given by
\begin{equation}\label{FC3}
\left\{ \begin{array}{l}
{\bar{v}^e_{k} = {(\, {\sum\limits_{m } {h_{mk}^2v_{m\to k}^{s{\;^{ - 1}}}(\tau)\,} } )^{ - 1}},}\\
\bar{x}_{k}^e = {(\bar{v}_k^l+\bar{v}^e_{k})}\sum\limits_{m } {h'_{mk}v_{m }^{s{\;^{ - 1}}}x_{m \to k}^s(\tau)}+\bar{x}_k^l,
\end{array} \right.\quad%\vspace{-0.1cm}
\end{equation}
where $k \in \mathcal{N}_u, m \in \mathcal{N}_s$. The calculation of the extrinsic estimation $\bar{x}_{k}^e$ is derived as follows.
\begin{equation}
\begin{gathered}
  \bar x_k^e \!=\! \bar v_k^e(\hat v_k^{ \!-\! 1}{{\hat x}_k} \!-\! \bar v_k^{{l^{ \!-\! 1}}}\bar x_k^l)
  %\quad  = \bar v_k^e\left( {(\bar v_k^{{e^{ - 1}}} \!\!+ \bar v_k^{{l^{ - 1}}})\bar v_k^l( {\sum\limits_m {{{h'}_{mk}}v_m^{s{\;^{ - 1}}}x_{m \to k}^s(\tau ) + \bar v_k^{{l^{ - 1}}}\bar x_k^l} } ) - \bar v_k^{{l^{ - 1}}}\bar x_k^l} \right) \hfill \\
    \!=\! {(\bar v_k^l \!+\! \bar v_k^e)\!\!\sum\limits_m {{{h}'_{mk}}v_m^{s{^{ \!-\! 1}}}x_{m \to k}^s(\tau )}  \!+\! \bar x_k^l}. \hfill \\
\end{gathered}%\vspace{-0.2cm}
\end{equation}
When the MSE of the SA-GMP meets the requirement, or the number of iterations reaches the limit, we output the \emph{a-posteriori} estimation $\hat{x}_{k}$ and its variance $\hat{v}_{k}$.
\begin{equation}\label{FC4}
\left\{ \begin{array}{l}
{ \hat{v}_{k} = {(\, {\sum\limits_{m } {h_{mk}^2v_{m \to k}^{s{\;^{ - 1}}}(\tau)\,} }+\bar{v}_{{k}}^{ l^{-1}} )^{ - 1}},}\\
\hat{x}_{k} = \bar{v}^l_{k}(\sum\limits_{m } {h'_{mk}\bar{v}_{m}^{s{\;^{ - 1}}}x_{m \to k}^s(\tau) + \bar{v}_{{k}}^{ l^{-1}}\bar{x}_k^{l}})\;,
\end{array} \right.
\end{equation}
for $k \in \mathcal{N}_u, m \in \mathcal{N}_s$. The detailed process of SA-GMP is given in Algorithm 2.
\begin{algorithm}[ht!]
\caption{SA-GMP Algorithm}
\begin{algorithmic}[1]
\State {\small{\textbf{Input:} {{$\bar{\mathbf{x}}^l$, $\bar{\mathbf{v}}^l_{\mathbf{x}}$, $\sigma^2_n$, $\epsilon>0$, $N_{ite}^{ese}$}}, $\mathbf{H}$, $\tilde{\gamma}$, calculate $w$, {{$\mathbf{H}^{(2)}$}}, $\mathbf{H}'$, $\textbf{{y}}'$, $\bar{\mathbf{v}}^s=[\bar{{v}}_1^s,\cdots,\bar{{v}}_{N_s}^s]^T$.
\State \textbf{Initialization:} {{$\tau\!=\!0$, $\mathbf{X}_{us}(0)\!=\!\bf{0}$}}, {{$\mathbf{V}_{us}(0)=+\boldsymbol{\infty}$.}}
\State \textbf{Do}
\State \quad\;\;$\tau\!=\!\tau\!+\!1$,
\State \quad\; $\widetilde{\mathbf{X}}_{us}(\tau)\!=\!\mathbf{X}_{us}(\tau).*\!\mathbf{H}'^T$ and $\widetilde{\mathbf{V}}_{us}(\tau)\!=\!\mathbf{V}_{us}(\tau).\!*{\mathbf{H}^{(2)}}^{T}$,
\State  \vspace{-0.35cm}\[ \!\!\!\!\begin{array}{l}
\left[\!\!\!\! \begin{array}{l}
{{\rm{\mathbf{X}}}_{su}}(\!\tau\!)\\
{\mathbf{V}_{su}}(\!\tau\!)
\end{array} \!\!\!\!\right]\!\! =\!\! \left[ \!\!\! \begin{array}{l}
\textbf{{y}}' \!\!-\! dia{g^{-1}}\!\{ \mathbf{1}\cdot{\widetilde{\mathbf{X}}_{us}}(\tau\!-\! 1)\} \\
\sigma_n^2\!\!\cdot\!\!{\mathbf{1}}\!\!+\!dia{g^{ \!- \!1}}\!\{ \mathbf{1}\!\cdot \!{\widetilde{\mathbf{V}}_{us}}(\tau \!-\! 1)\!\}
\end{array} \!\!\!\!\right] \!\!\cdot\! \!{\mathbf{1}} \!\!-\! (w\!-\!1)\!\!\left[ \!\!\! \begin{array}{l}
 {{\rm{\mathbf{X}}}_{su}}(\!\tau\!-\!1)\\
\qquad\mathbf{0}
\end{array} \!\!\!\!\right]\!\!,
\end{array}\]}}
\State\vspace{-0.55cm}\[ \widetilde{\mathbf{W}}_{su}(\!\tau\!)\!=\!\mathbf{H}^{(2)}.*\mathbf{V}^{(\!-\!1\!)}_{su}(\!\tau\!), \widetilde{\mathbf{G}}_{su}(\!\tau\!)\!=\!\mathbf{H}'.*diag\{\bar{\mathbf{v}}^s\}.*\mathbf{X}_{su}(\!\tau\!),\]
\State \vspace{-0.6cm}\[\;\: \begin{array}{l}
\left[ \!\!\!\!\begin{array}{l}
{\mathbf{W}_{us}}(\!\tau\!)\\
{\mathbf{G}_{us}}(\!\tau\!)
\end{array} \!\!\!\!\right]\! \!=\!\! \left[\!\! \!\!\begin{array}{l}
\bar{\mathbf{v}}^{l^{(\!-\!1\!)}}_{\mathbf{x}} \!\!+\! dia{g^{ \!-\! 1}}\!\left\{ {{\mathbf{1}} \cdot {{\widetilde{\mathbf{W}}_{su}}(\tau)}} \right\}\mathop {}\limits_{\mathop {}\limits_{} }\\
 \bar{\mathbf{v}}^{l^{(\!-\!1\!)}}_{\mathbf{x}}\!\!.\!*\!\bar{\mathbf{x}}^l \!\!+\! dia{g^{ \!-\! 1}}\!\left\{ \!{{\mathbf{1}} \!\cdot\! \widetilde{\mathbf{G}}_{su}(\!\tau\!)} \!\right\}
\end{array}\!\!\! \!\!\right]\!\! \cdot\!\! {\mathbf{1}}\!-\!\!\left[\!\!\! \begin{array}{l}
\widetilde{\mathbf{{W}}}_{su}^T(\!\tau\!)\\
\widetilde{\mathbf{G}}_{su}^T(\!\tau\!)
\end{array}\!\!\! \right]\!\!,
\end{array}\qquad\qquad\]\vspace{-0.2cm}
\vspace{0cm}
\State \quad$\mathbf{V}_{us}(\!\tau\!)\!=\!\mathbf{W}_{us}^{(-1)}(\!\tau\!)$ and  $\mathbf{X}_{us}(\!\tau\!)\!=\!diag\{\bar{\mathbf{v}}^l\}.*\mathbf{G}_{us}(\!\tau\!)$.
\State \textbf{While} \;{\small{$\left( \:|\mathbf{E}_{us}(\tau)-\mathbf{E}_{us}{(\tau-1)}|>\epsilon \;{\textbf{or}}\; \tau\leq N_{ite}^{ese} \;\right)$}}
\State\vspace{-0.5cm} {{ \[  \begin{array}{l}
\bar{\mathbf{v}}^e = {\left( {dia{g^{ - 1}}\left\{ {{\mathbf{1}_{K \times M}} \cdot {\widetilde{\mathbf{W}}_{su}}(\tau) } \right\}} \right)^{( - 1)}}\!\!\!\!,\\
\bar{{\mathbf{x}}}^e = (\bar{\mathbf{v}}^l + \bar{{\mathbf{v}}}^e).*dia{g^{ - 1}}\left\{ {\mathbf{1}_{K \times M}} \cdot \widetilde{\mathbf{G}}_{su}(\tau) \right\}+\bar{{\mathbf{x}}}^l.
\end{array} \qquad\qquad\qquad\qquad\]}}\vspace{-0.2cm}
\State \textbf{Output:}  $\bar{{\mathbf{x}}}^e$ and $\bar{\mathbf{v}}^e $ .
\end{algorithmic}
\end{algorithm}

\textbf{\emph{Remark 4}}: The linear operators: scaling $\mathbf{y}$ and $\mathbf{H}$ to ${\mathbf{H}'}=\sqrt{w}\mathbf{H}$ and $\textbf{{y}}'=\sqrt{w}\textbf{{y}}$ with a relaxation parameter $w$, and adding a new term $\!(\!w\!-\!1\!)x_{m \to k}^s\!(\!\tau\!-\!1\!)$ for the mean message update at the SN, are aimed at minimizing the spectral radius and assuring the convergence of SA-GMP (see Theorem 3 and Corollary 4). Based on Theorem 1, the variances of GMP converge to that of the optimal LMMSE detection. Therefore, we keep the variance update of SA-GMP the same as that of the GMP. From Theorem 2, we can see that the GMP converges to the LMMSE if it is convergent and $\hat{v}=\bar{v}^l$. Therefore, we let $v_{k\to m}^u(\tau)=\bar{v}_k^l$ and $v_{m\to k}^s(\tau)=\bar{v}_m^s$ in the mean message update (\ref{FC1}), to assure that the SA-GMP converges to the LMMSE detection.
\newcommand{\tabincell}[2]{\begin{tabular}{@{}#1@{}}#2\end{tabular}}
\begin{table*}[b!]
\renewcommand{\arraystretch}{1.6}
\centering
\caption{Complexity comparison of the different detections for MIMO-NOMA system}
\label{table2}
\begin{tabular}{|c|c|c|}
\hline
 \tabincell{c}{NONITERATIVE\vspace{-0.05cm}\\ DETECTION ALGORITHMS} & IF/ZF  & LMMSE \\
\hline
  COMPLEXITY & $\mathcal{O}\left(\,(N_s^2N_u+N_s^3)N_{ite}^{out}\,\right)$ &  $\mathcal{O}\left(\,\min\{N_sN_u^2+N_u^3,\;N_uN_s^2+N_s^3\}N_{ite}^{out}\,\right)$\\
\hline \hline
\tabincell{c}{ITERATIVE\vspace{-0.05cm}\\DETECTION ALGORITHMS}  & Jacobi\;  \& GaBP  \& Richardson& \tabincell{c}{{AMP} \& GMP\& SA-GMP }\\
\hline
  COMPLEXITY& $\mathcal{O}\left(\,N_sN_u^2 + N_u^2N_{ite}^{ese}N_{ite}^{out}\,\right)$\!\!&$\mathcal{O}(\,{N_uN_s}N_{ite}^{ese}N_{ite}^{out}\,)$  \\
\hline
\end{tabular}
\end{table*}
\subsection{Variance Convergence of SA-GMP}
The variance convergence of SA-GMP can be derived by the SE in \cite{Donoho2010a,Donoho2009,Bayati2011}. As the variance update of SA-GMP is the same as that of the GMP, the variances of SA-GMP converge to the same values as those of the GMP, i.e., they converge to the exact MSE of the LMMSE multi-user detection given in (\ref{p42}) and (\ref{p8}).

\subsection{Mean Convergence of SA-GMP}
Similar to the analysis in GMP, for the symmetric overloaded massive MIMO-NOMA system, we have ${\bar v}_k^l={\bar v}^l$ and ${\bar v}_m^s={v}^s$, for $k \in \mathcal{N}_u, m \in \mathcal{N}_s$.  To simplify the expressions, we define two parameters like that in the GMP as following.
\begin{equation}\label{FCpara}
\tilde{\theta} \! = \!\frac{{{{\bar v}^l}}}{{\sigma _n^2}}\! =\! snr,\;
\tilde{\gamma}  \!=\! \frac{{\bar v}^l}{{{v^s}}} \!=\! \frac{{{{\bar v}^l}}}{{{N_u}{{\bar v}^l} \!+\! \sigma _n^2}} \!=\! {({N_u} \!+\! sn{r^{ - 1}})^{ - 1}}.
\end{equation}

\subsubsection{Mean Convergence of SA-GMP}
We obtain the following lemmas for the mean convergence of SA-GMP.

\textbf{\textit{Lemma 2:}} \emph{In the overloaded massive MIMO-NOMA with $\mathbf{V}_{\bar{\mathbf{x}}}^l=\bar{v}^lI_{N_u}$, the sum-node messages of satisfy $x_{ m\to k}^s(\tau)=x^s_{m}(\tau)$, $\forall m \in \mathcal{N}_s$. The SA-GMP converges to the following classical iterative algorithm.
\begin{equation}
\mathbf{x}^s(\!\tau\!) \!=\! \textbf{{y}}' \!-\!  \left[\tilde{\gamma} \mathbf{H}'\mathbf{H}'^T \!+\!\big((1\!-\!\tilde{\gamma}N_u )w\!-\!1\big)\mathbf{I}_{N_s}\!\right] \!\mathbf{x}^s(\tau-1)-\! \mathbf{H}'\bar{\mathbf{x}}^l.
\end{equation}}
\begin{IEEEproof}
See APPENDIX E.
\end{IEEEproof}

Based on Lemma 2 and Proposition 3, we have the following theorem.

\textbf{\emph{Theorem 3:}}  \emph{In the overloaded massive MIMO-NOMA, where $\beta= N_u/N_s$ is fixed, $N_u$ is large, and $\mathbf{V}_{\bar{\mathbf{x}}}^l=\bar{v}^lI_{N_u}$, the SA-GMP converges to the LMMSE estimation if $w$ satisfies $0<w<2/\lambda_{max}^\mathbf{A}$, where $\lambda_{max}^\mathbf{A}$ is the largest eigenvalue of matrix $\mathbf{A}= \tilde{\gamma}\mathbf{H}\mathbf{H}^T +(1-\tilde{\gamma}N_u )\mathbf{I}_r$.}
\begin{IEEEproof}
See APPENDIX F.
\end{IEEEproof}
\begin{table*}[t]
\renewcommand{\arraystretch}{1.4}
\caption{Convergence comparison between SA-GMP, AMP, GMP, Jacobi, GaBP and Richardson algorithms, where $``+"$ and $``-"$ denote the right and left limit respectively.}
\label{table}
\centering
\begin{tabular}{|c|c|c|c|c|c|}
\hline
 $\beta$ & \tabincell{c}{ Jacobi  \& GaBP} &  \tabincell{c}{GMP}  & {AMP}  & \tabincell{c}{Richardson \& SA-GMP} \\
\hline
$\beta>(\sqrt{2}-1)^{-2}$ & Converge to LMMSE& \tabincell{c}{Converge, not\vspace{-0.05cm}\\ approach LMMSE} & {Converge to LMMSE} & Converge to LMMSE \\
\hline
$\beta\to(\sqrt{2}-1)^{-2}_-$ & Diverge & \tabincell{c}{Converge, not\vspace{-0.05cm}\\ approach LMMSE} & {Converge to LMMSE} & Converge to LMMSE\\
\hline
$\beta\to1_+$ & Diverge &Diverge & {Diverge} & Converge to LMMSE\\
\hline
\end{tabular}
\end{table*}
\subsubsection{Spectral Radius Minimization}
{The spectral radius} in SA-GMP can be minimized by the following corollary.

{\textbf{\emph{Corollary 3:}} \emph{The optimal relaxation parameter is
  \begin{equation}
 w^* =1/(1+\tilde{\gamma} N_s),
 \end{equation}
which minimizes {the spectral radius} as follows.
 \begin{equation}
 \rho_{_{SA-GMP}} = \frac{2\tilde{\gamma}\sqrt{N_uN_s}}{1+\tilde{\gamma} N_s}<1.
 \end{equation}
In addition, we have
 \begin{equation}
  {\rho_{_{SA-GMP}}} <{\rho_{_{GMP}}}.
 \end{equation}
}%\vspace{0.25cm}
\begin{IEEEproof}
See APPENDIX G.
\end{IEEEproof}}

\subsubsection{Comparison with GMP}
The SA-GMP converges to LMMSE with a rate $\rho_{_{SA-GMP}}^{\tau}$. In addition, the spectral radius of the proposed SA-GMP is strictly smaller than that of the GMP, which means the SA-GMP has a faster convergence speed. Hence, we have the following corollary.\vspace{0.25cm}

\textbf{\emph{Corollary 4:}} \emph{In the overloaded massive MIMO-NOMA system, where $\beta= N_u/N_s$ is fixed, $N_u$ is large, and $\mathbf{V}_{\bar{\mathbf{x}}}^l=\bar{v}^lI_{N_u}$, the proposed SA-GMP converges faster than the GMP.}\vspace{0.25cm}

{\textbf{\textit{Remark 5}}: The proof of Theorem 3 and the spectral radius analysis contain the detailed process for the SA-GMP design. The reason why we propose the SA-GMP is that its spectral radius is minimized after the linear modifications. As a result, the convergence prerequisite and convergence speed of the SA-GMP are significantly improved.}

\subsection{Complexity Comparison of Different Detectors}
Table \ref{table2} compares the complexity of the GMP and that of other detections. The complexities of Zero Forcing (ZF) or Inverse Filter (IF) are both $\mathcal{O}\left((N_s^2N_u+N_s^3)N_{ite}^{out}\right)$, where $\mathcal{O}(N_s^3)$ arises from the matrix inversion, $\mathcal{O}(N_u N_s^2)$ arises from the calculation of $\mathbf{H}^T\mathbf{H}$ and $N^{out}_{ite}$ denotes the number of outer iterations between the ESE and decoders. The complexity of the LMMSE detector is $\mathcal{O}\left(\,\min\{N_sN_u^2+N_u^3, \;N_u N_s^2+N_s^3\}N_{ite}^{out}\,\right)$, which is mentioned in section II-B. Similarly, the complexity in each ESE iteration of the above classical iterative algorithms is $O(N_u^2)$, and the matrix calculation $\mathbf{A} = \mathbf{I}_{N_u} + snr\mathbf{H}^T\mathbf{H}$ costs $O(N_s N_u^2)$ operations before the ESE iteration. Hence, the total complexity of the classical iterative algorithm is $O(N_sN_u^2+N_u^2N_{ite}^{ese}N_{ite}^{out})$. The complexity of AMP algorithm is $\mathcal{O}(N_uN_sN_{ite}^{ese}N_{ite}^{out})$.

It should be pointed out that the complexities of GMP and SA-GMP algorithms are the same, i.e., $\mathcal{O}(N_uN_sN_{ite}^{ese}N_{ite}^{out})$ (see Section III. E). Besides, if the channel is sparse, the complexity of SA-GMP (or GMP) can be further reduced to $\mathcal{O}(N_\mathbf{H}N_{ite}^{ese}N_{ite}^{out})$, where $N_\mathbf{H}$ is the number of nonzero elements in channel matrix $\mathbf{H}$. However, the complexities of IF/ZF, LMMSE, Jacobi, GaBP and Richardson algorithms will not change by the sparsity of channel matrix, because the sparse property is destroyed after calculating $\mathbf{H}^T\mathbf{H}$. In general, we often let $N_{ite}^{ese}=1$ and consider only the iteration between the ESE and decoders. In this case, the complexity of GMP or SA-GMP is only $1/N_u$ (or $1/N_s$) of the other conventional detectors. %\vspace{-0.2cm} %\vspace{-0.2cm}

\section{Simulation Results}
In this section, we present the numerical results of the proposed detectors for the overloaded massive MIMO-NOMA system. We assume that the sources are i.i.d. with $\mathcal{N}^{N_u}(0,1)$ and the entries of the channel matrix $\mathbf{H}$ are i.i.d. with $\mathcal{N}^{N_s\times N_u}(0,1)$. In the following simulations, $SNR=\dfrac{1}{\sigma^2_n}$ is the signal-to-noise ratio, and $MSE=\frac{1}{N_u}\cdot{\mathbf{E}}\left[\|\mathbf{x}-\hat{\mathbf{x}}\|^2_2\right]$ denotes the averaged MSE of the estimation. All the simulations are repeated with 500 random realizations.

\begin{figure}[t]
  \centering
  \includegraphics[width=7.0cm]{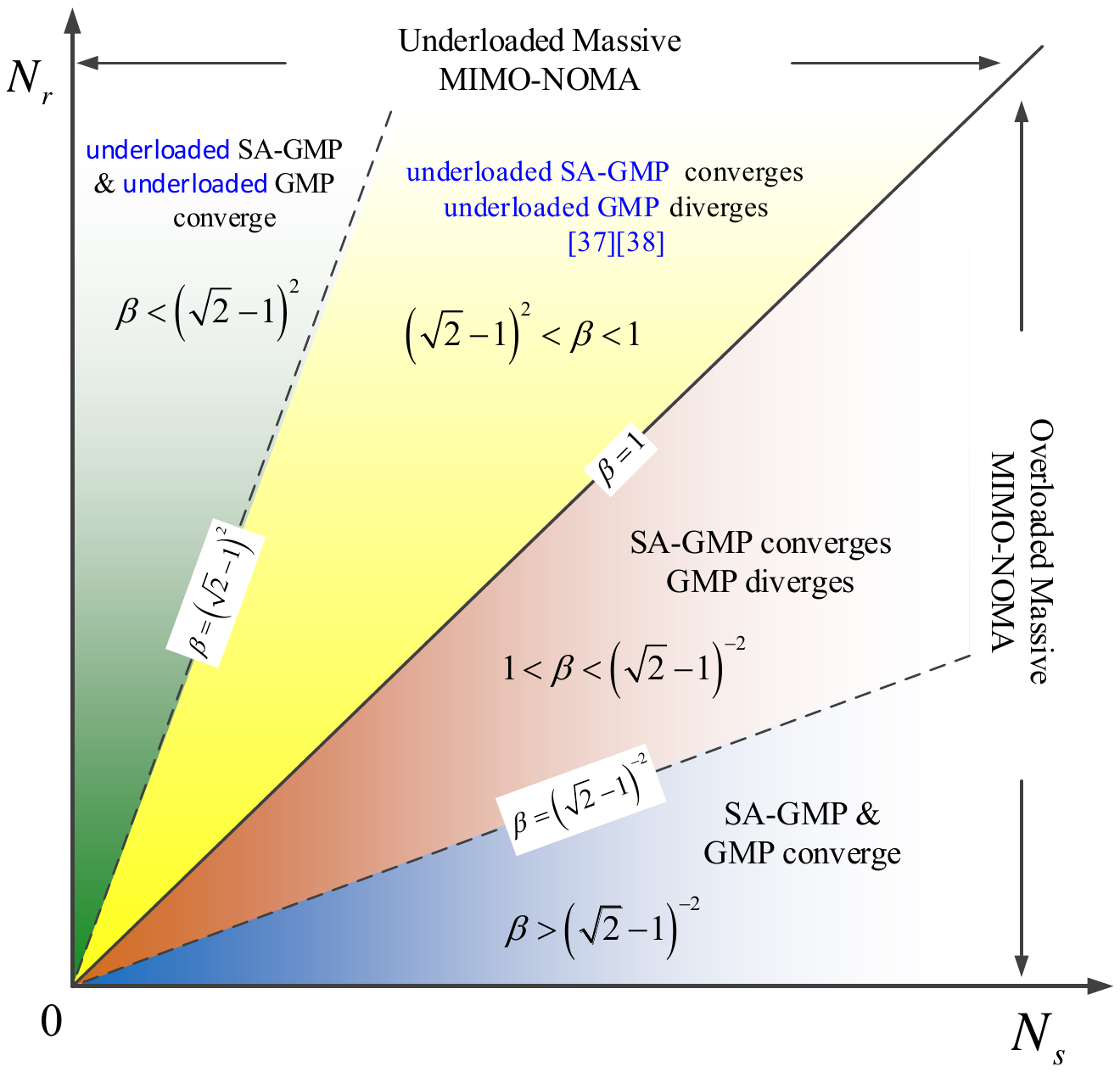}\\
  \caption{Convergence region of GMP and SA-GMP: GMP diverges when $(\sqrt{2}-1)^2<\beta<(\sqrt{2}-1)^{-2}$, and SA-GMP converges for any value of $\beta$. The underloaded case was discussed in \cite{Lei2016}.} \label{conv_region}
\end{figure}
\subsection{Convergence Comparison}
Table \ref{table} concludes the convergence comparison of the different overloaded massive MIMO-NOMA detections, where $``+"$ (or $``-"$) denotes right limit (left limit). It shows that: 1) all the iterative algorithms are convergent when $\beta>(\sqrt{2}-1)^{-2}$; 2) the Jacobi and GaBP algorithms are divergent when $\beta<(\sqrt{2}-1)^{-2}$; 3) AMP, GMP and SA-GMP are still convergent when $\beta$ is close to $(\sqrt{2}-1)^{-2}_-$; 4) Richardson algorithm and SA-GMP are convergent even when $\beta$ is close to $1$. {Fig. \ref{conv_region} shows the convergence region of the GMP and SA-GMP. The GMP only converges} {when either $\beta<(\sqrt{2}-1)^{2}$ or $\beta>(\sqrt{2}-1)^{-2}$, but diverges when $(\sqrt{2}-1)^2<\beta<(\sqrt{2}-1)^{-2}$. However, the SA-GMP converges for any value of $\beta$.}
\begin{figure}[t]
  \centering
  \includegraphics[width=7.5cm]{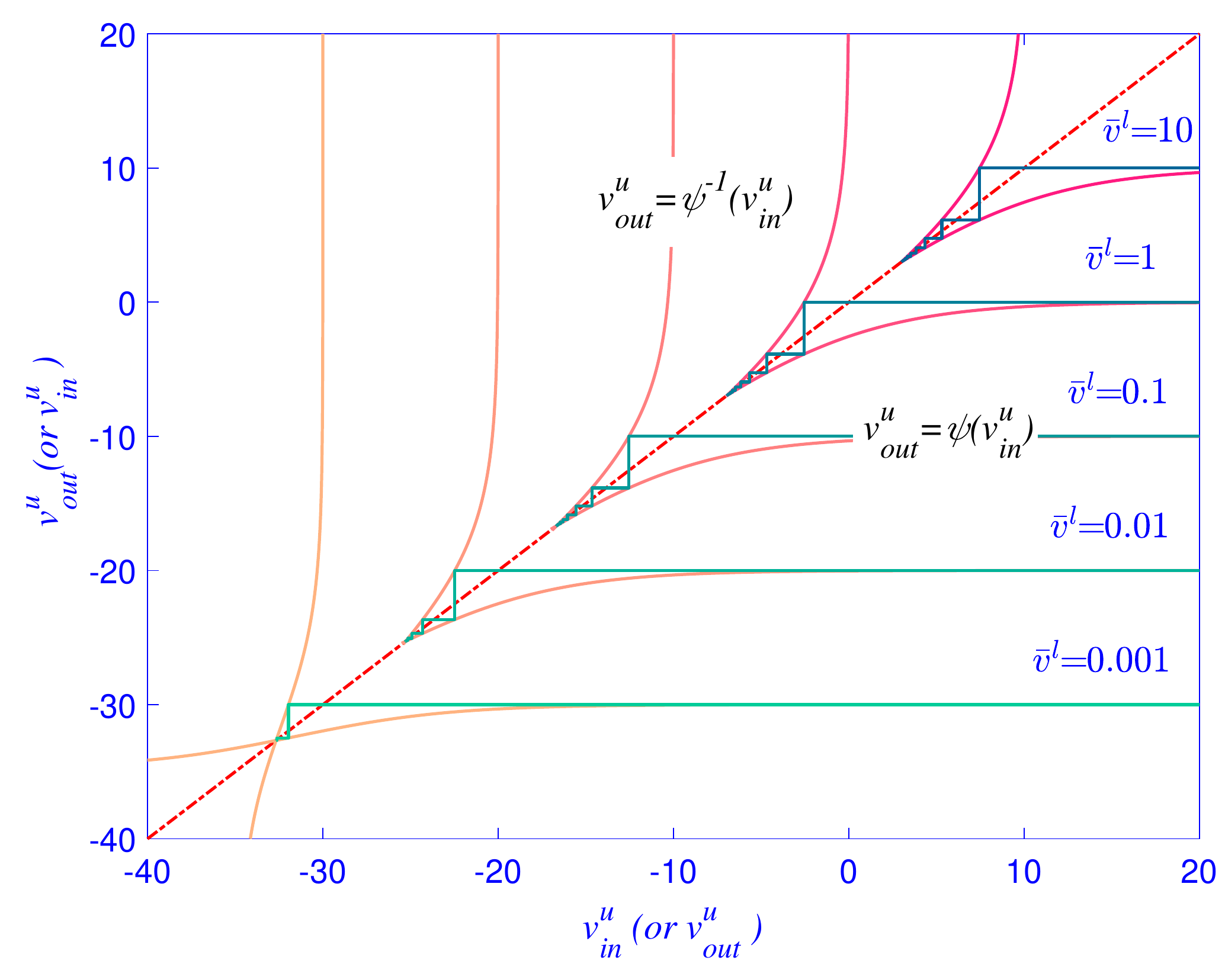}\\
  \caption{MSET chart analysis of GMP, where $N_u\!=\!250$, $N_s\!=\!200$, $SNR\!=\!20$(dB), $\bar{v}^l\!=\![10, \;1, \; 0.1,\; 10^{-2}, \;10^{-3}]$. $\bar{v}^l$ is the input variance of the estimation from decoders, and $v^u_{in}$ and $v^u_{out}$ are the input and output variance of the message of GMP at each iteration. The estimated variance decreases with the increasing of iterations, and decreases with the decreasing of $\bar{v}^l$.
}\label{f4}
\end{figure}

\subsection{MSET of GMP}
{In Fig. \ref{f4}, we track the input and output MSEs (or the input and output variances: $v^u_{in}$ and $v^u_{out}$) during the iteration of the GMP under different input $\bar{v}^l$. In this simulation, we consider the cases that $N_u=250$, $N_s=200$, $SNR=20$(dB) and $\bar{v}^l=[10, \;1, \; 0.1,\; 10^{-2}, \;10^{-3}]$. The ``MSE Transfer (MSET) chart" is employed to analyse the system performance, which is analogous to the EXIT chart \cite{Bhattad2007}. In fact, the MSET chart can be transformed into the EXIT chart based on the relationship between the mutual information and MSE \cite{Bhattad2007, Guo2005}.} Fig. \ref{f4} shows that for the uncoded overloaded massive MIMO-NOMA, the MSE always converges to a high MSE. However, for the coded overloaded massive MIMO-NOMA, the fixed point decreases with the decreasing of $\bar{v}^l$, which means that the decoders improve the MSE. Furthermore, in the overloaded massive MIMO-NOMA, the GMP converges very fast to the fix point (with less than 10 iterations). Furthermore, the convergence speed increases with the decreasing of $\bar{v}^l$. %\vspace{-1.2cm}
\begin{figure}[t]
  \centering
  \includegraphics[width=9.0cm]{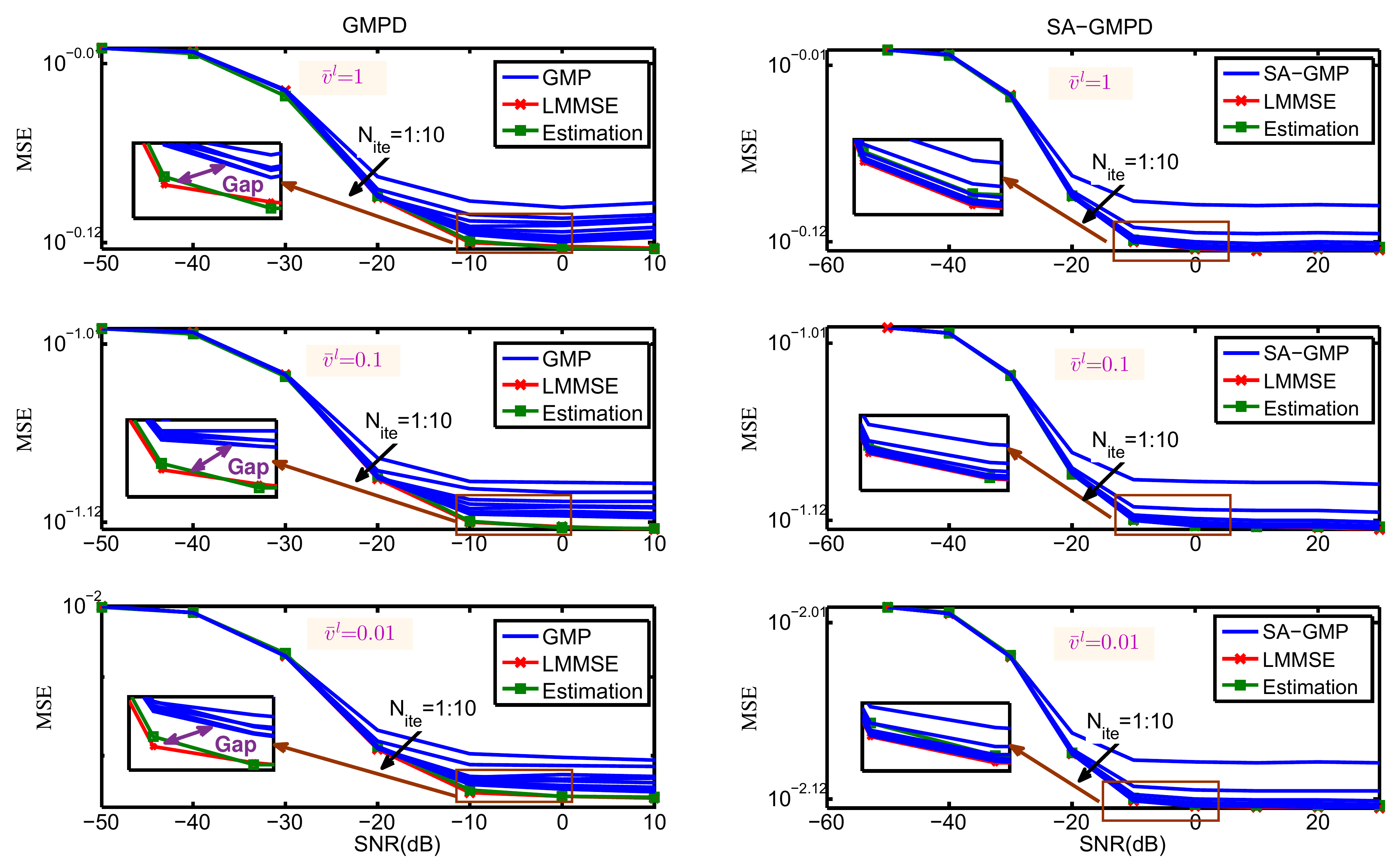}\\
  \caption{MSE comparison between the LMMSE detection, GMP and SA-GMP with $1 - 10$ iterations; the performance estimate of the LMMSE detection and the proposed GMP. $N_u\!=\!400$, $N_s\!=\!100$, $\beta=4$ and $\bar{v}^l\!=\![1, \; 0.1,\; 10^{-2}]$. The proposed SA-GMP converges to the LMMSE detection, and the GMP converges to a higher MSE than that of the LMMSE detection. }\label{f5}
\end{figure}
\subsection{MSE Comparison between LMMSE, GMP, and SA-GMP}
Fig. \ref{f5} presents the average MSE comparison between the LMMSE detection, GMP and SA-GMP when the iteration number increased from 1 to 10, where $N_u=400$, $N_s=100$, $\beta=4$ and $\bar{v}^l=[1, 0.1, 10^{-1}, 10^{-2}, 10^{-3}]$. The performance estimate (\ref{PA6}) is also shown in Fig. \ref{f5}. It is observed that under the different input $\bar{v}^l$, the performance estimate of the LMMSE detection (\ref{PA6}) is accurate. From the three subfigures in Fig. \ref{f5} on the left, we can see that the GMP converges to a higher MSE than that of the LMMSE detection for any input $\bar{v}^l$, which coincides with the conclusions in Theorem 2 and its corollaries. However, from the three subfigures in Fig. \ref{f5} on the right, it can be seen that the proposed SA-GMP always converges to the LMMSE detection and has a faster convergence speed than the GMP, which verifies the conclusions in Theorem 3 and its corollaries. Furthermore, these results also agree with the MSET charts in Fig. \ref{f4}.

\begin{figure}[t]
  \centering
  \includegraphics[width=8.9cm]{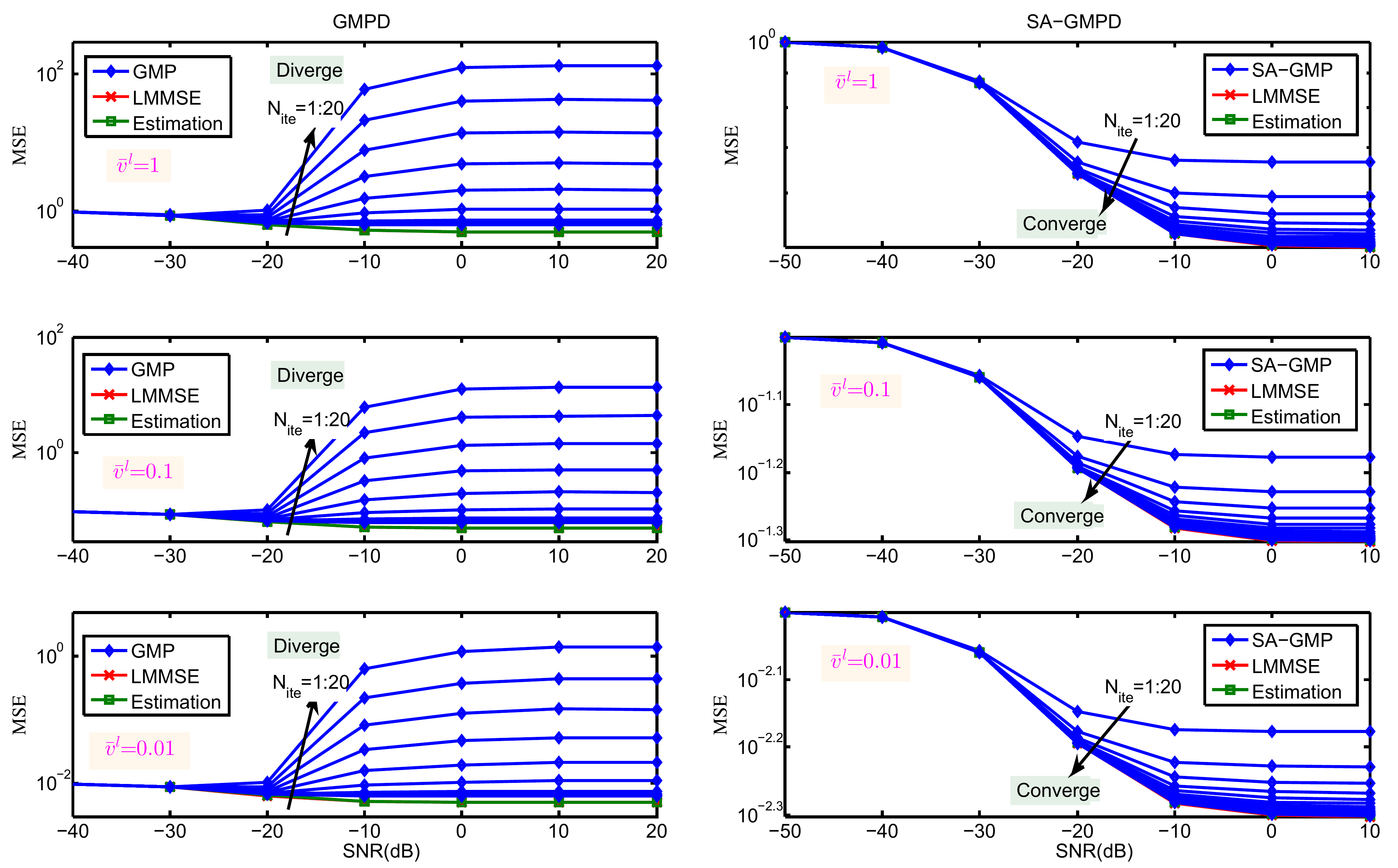}\\
  \caption{MSE of the SA-GMP converges to that of the LMMSE detection with $1 - 20$ iterations. However, the MSE of GMP diverges in this case. $N_u=400$, $N_s=200$, $\beta=2$, $\bar{v}^l=[1, \; 0.1,\; 10^{-2}]$. }\label{f6}
\end{figure}

Fig. \ref{f6} gives the average MSEs of SA-GMP and GMP for the cases that $N_u=400$, $N_s=200$, $\beta=2$ and $\bar{v}^l=[1, \; 0.1,\; 10^{-2}]$. It can be seen that the proposed SA-GMP always converges to the LMMSE detection with a fast convergence speed for any input $\bar{v}^l$ and for any small value of $\beta$. However, the MSE of GMP diverges for this small value of $\beta$. This verifies the analysis result in Theorem 3.

\subsection{Comparison with AMP}
\begin{figure}[t]
\centering
\begin{tabular}{cc}
\includegraphics[width=.22\textwidth]{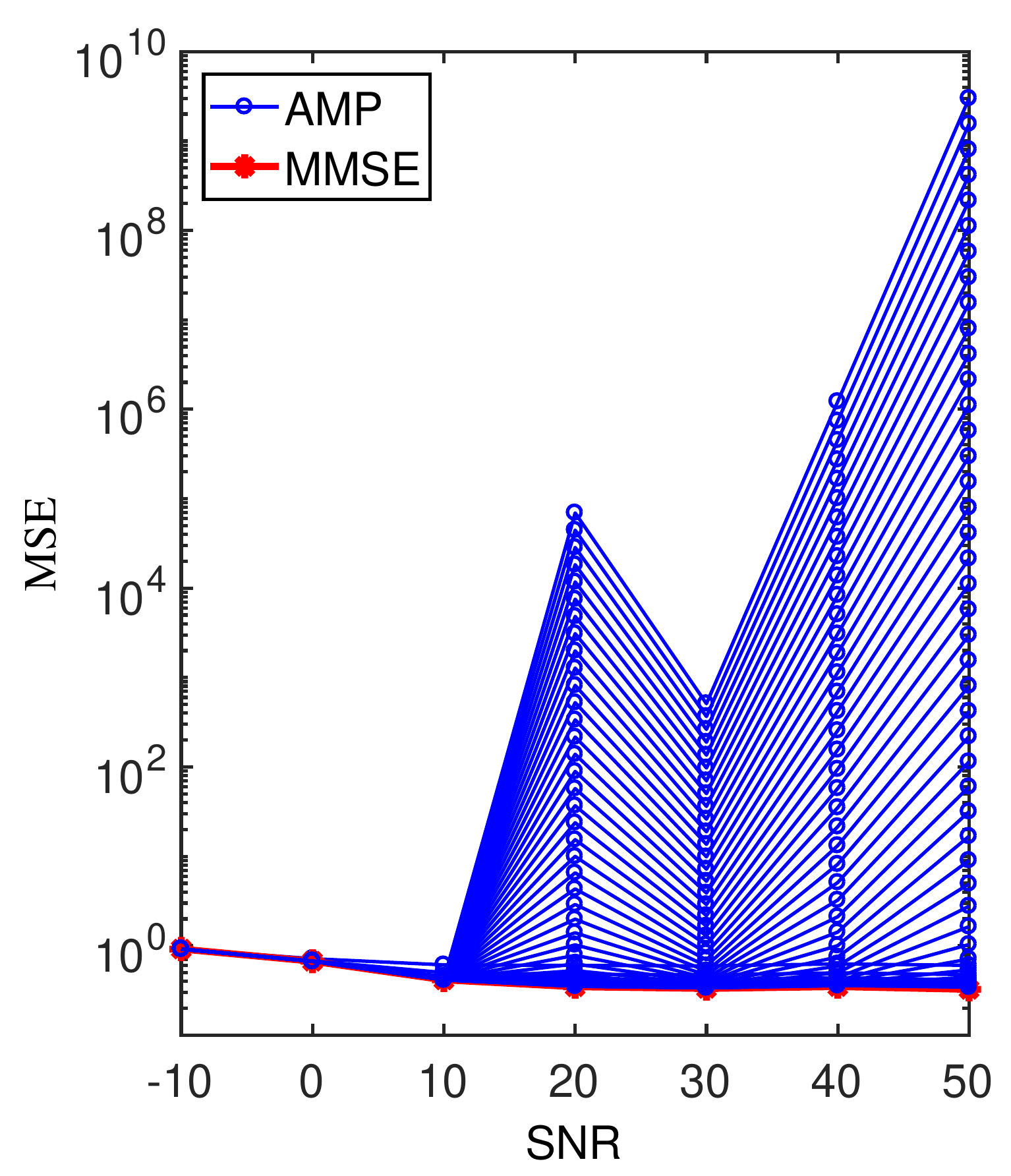}          &
\includegraphics[width=.22\textwidth]{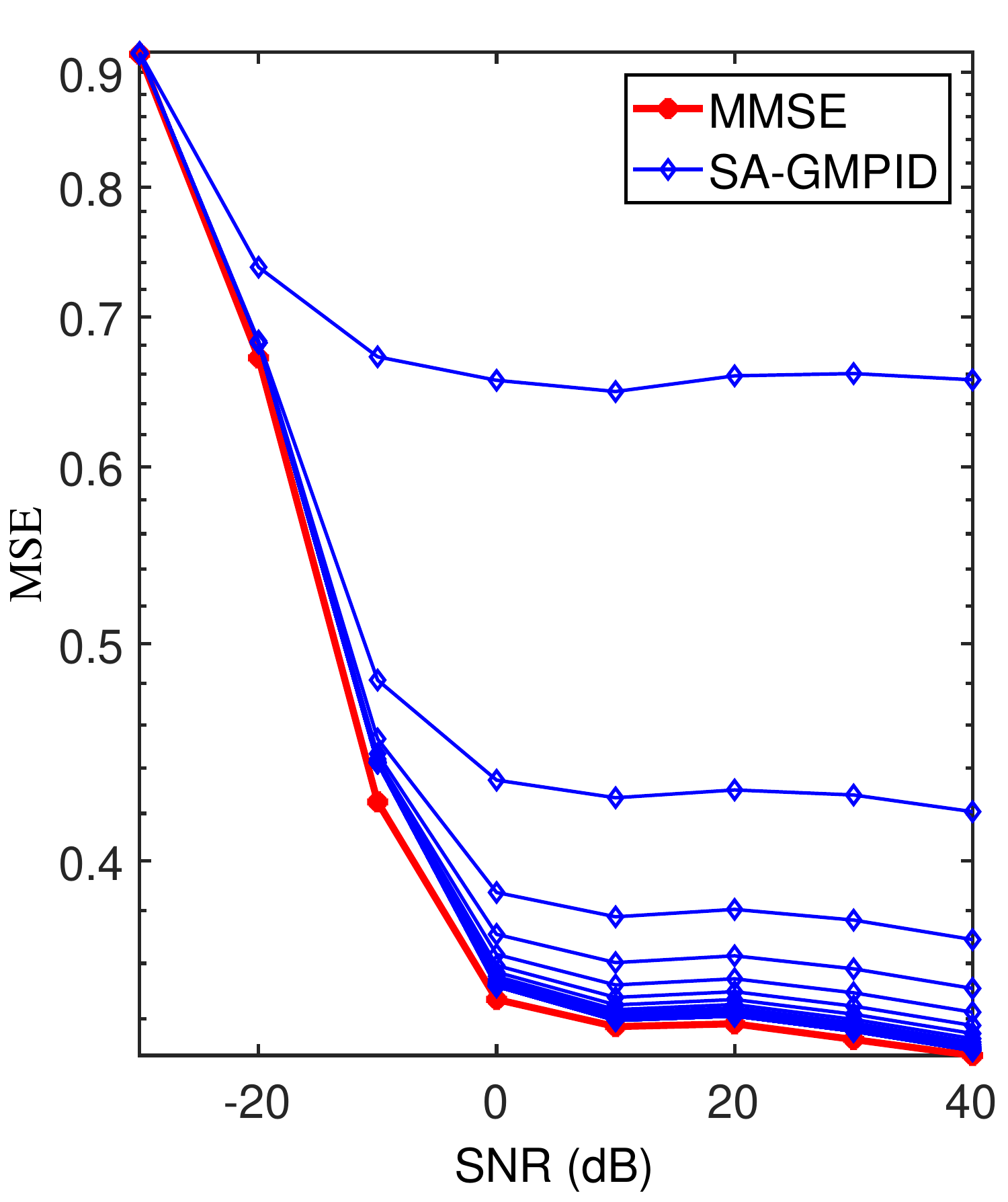}\\
$\quad$\footnotesize{(a) AMP}  & $\quad$\footnotesize{(b) SA-GMP}
\end{tabular}
\caption{MSE of the SA-GMP converges to that of the LMMSE detection with $50$ iterations. However, the MSE of AMP diverges in these cases. In (a) and (b), $N_u=150$, $N_s=100$, $\beta=3/2$.}
\label{fig_AMP}
\end{figure}

To compare with the AMP algorithm\footnote{\emph{\textbf{Note 1:}} In \cite{Bayati2011,Donoho2009}, the measurement matrix satisfies $\mathbf{A}\sim\mathcal{N}^{N_s\times N_u}(0,1/N_s)$, but in this paper we assume $\mathbf{H}\sim\mathcal{N}^{N_s\times N_u}(0,1)$. Hence, in order to compare with the AMP detection, the system model is rewritten to $\mathbf{y}'=(\frac{1}{\sqrt{N_s}}\mathbf{H})\cdot\mathbf{x}+\mathbf{n}/\sqrt{N_s} =\mathbf{A}\mathbf{x}'+\mathbf{n}'$, where $\mathbf{n}'\sim \mathcal{N}^{N_s}(0,1/N_s)$. The AMP detection in \cite{Donoho2009} and \cite{Bayati2011} is used to recover $\hat{{\mathbf{x}}}$ from $\mathbf{y}'$.

\emph{\textbf{Note 2:}} Actually, even for $\mathbf{y} =\mathbf{A}\mathbf{x}+\mathbf{n}$, where $\mathbf{A}\sim\mathcal{N}^{N_s\times N_u}(0,1/N_s)$ and $\mathbf{x}\sim \mathcal{N}^{N_u}(0,1)$, when the system load equals to 3/2, the AMP algorithm is also divergent.}, we employe the AMP in \cite{Bayati2011} (in Section II.A) for the MIMO-NOMA detection.
It should be noted that when the system load satisfies $\beta>>1$ (e.g. $\beta= 6$), the AMP can converge to the optimal LMMSE detection. However, as shown in Fig. \ref{fig_AMP}, when the system load is close to 1 (e.g. $\beta= 3/2$ with $N_u=150$ and $N_s=100$), the AMP diverges, while the proposed SA-GMP always converges to the optimal LMMSE detection. The results of having AMP diverges is similarly applied to other algorithms e.g. GaBP.  {The reason is that AMP may become unreliable in certain settings, such as in ill conditioned channels\footnote{The condition number of $\mathbf{H}$ is defined as
$\kappa(\mathbf{H}) = \sqrt{\frac{\lambda_{\mathrm{max}}(\mathbf{H}\mathbf{H}^T)} {\lambda_{\mathrm{min}}(\mathbf{H}\mathbf{H}^T)}}$, where ${\lambda}_{max}(\mathbf{H}\mathbf{H}^T)$ and ${\lambda}_{min}(\mathbf{H}\mathbf{H}^T)$ denote the largest and smallest eigenvalues of matrix $\mathbf{H}\mathbf{H}^T$ respectively. The condition number can be estimated by \emph{random matrix theory}, such as $\kappa(\mathbf{H}) =  \frac{1+\beta^{-1}}{1-\beta^{-1}}$
where $\beta=N_u/N_s$. If $\beta\to1$, $\kappa(\mathbf{H})\to\infty$, i.e. the channel is ill conditioned. If $\beta\gg1$, $\kappa(\mathbf{H})\to1$, i.e. the channel is well conditioned.} \cite{Samuel2017, Ma2017}.} Apart from that, for the non-sparse signal recovery, the AMP is actually equivalent to the classic iterative algorithm. Hence, the AMP diverges when the spectral radius is larger than 1 (or $\beta$ closes to 1). Specifically, from Eq. (2.7) in \cite{Bayati2011}, we can see that the AMP diverges if $\tau_{t+1}^2>\tau_{t}^2$ (or in other words, the AMP can not always satisfy $\tau_{t+1}^2<\tau_{t}^2$ for the overloaded MIMO-NOMA system).
\begin{figure}[b!]
  \centering
  \includegraphics[width=7.5cm]{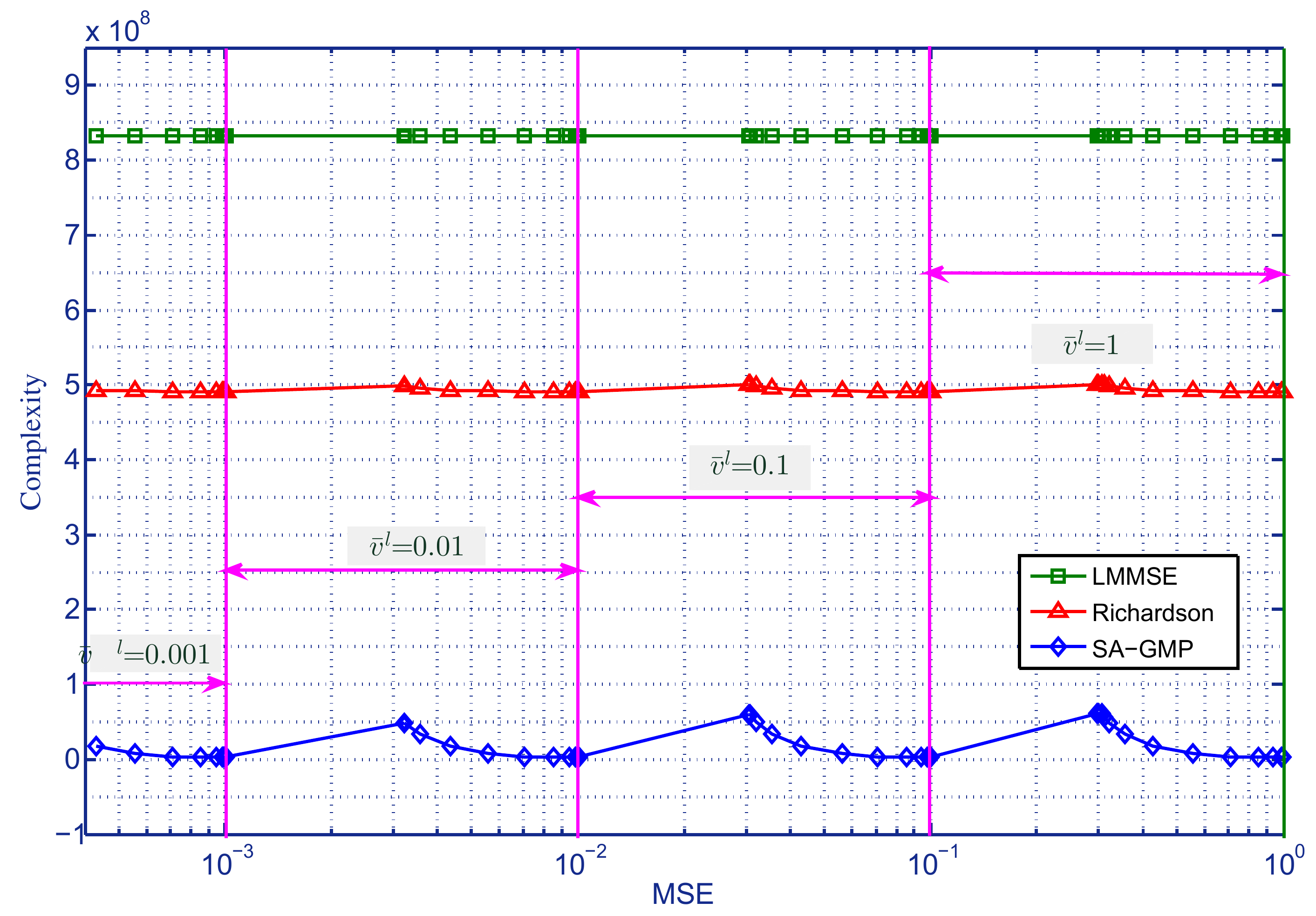}\\
  \caption{Complexity comparison between the LMMSE detection, Richardson algorithm and SA-GMP for $N_u=1000$, $N_s=700$, $\beta=10/7$ (close to 1), and $\bar{v}^l=[1, \; 0.1,\; 10^{-2},\; 10^{-3}]$. For any input $\bar{v}^l$, the LMMSE detection has a fixed and highest complexity, and the proposed SA-GMP has the lowest complexity. }\label{f7}
\end{figure}
\begin{figure*}[b!]
  \centering
  \includegraphics[width=12.0cm]{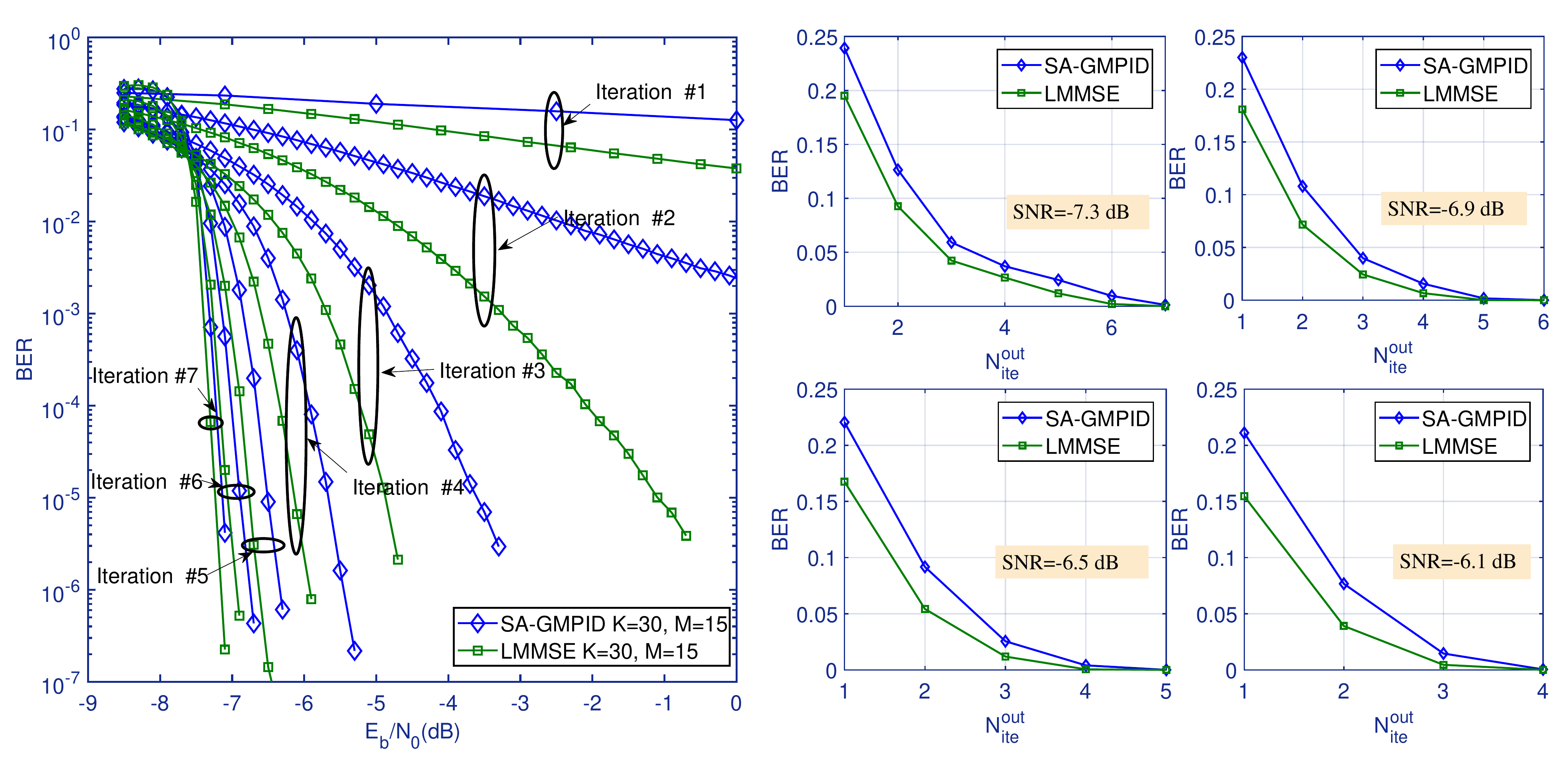}\\
  \caption{BER performances of SA-GMP and LMMSE detection with Gaussian approximation for the discrete overloaded MU-MIMO system, where $N_u=30$, $N_s=15$, $\beta=2$ and $N_{ite}^{ese}=1$, $N_{ite}^{out}=1 - 7$. Each user is encoded by a Turbo Hadamard code (Hadamard order 5, component codes 3, spread length 4) with code rate 0.01452 bits/symbol and code length $2.82\times 10^5$, and then 10bits superposition coded modulation is employed for the Turbo Hadamard code. Transmitting length of each user is $2.82\times 10^4$. The rate of each user is 0.1452 bits/symbol, and the system sum rate is 4.356 bits per channel use.}\label{f8}
\end{figure*}

\subsection{Complexity Comparison}
Fig. \ref{f7} illustrates the computational complexity comparison between the different detections under the different MSEs for the $1000\times700$ ($\beta=10/7$ ) overloaded massive MIMO-NOMA under the different input $\bar{v}^l$. The relative MSE error of each iterative algorithm is less than 0.05. We can see that the complexity of the SA-GMP detector increases with decreasing MSE, i.e., the lower MSE we need, the higher complexity it costs. It also shows that 1) the LMMSE detector has a constant but the highest complexity, 2) the classical iterative algorithm like Richardson algorithm always have much higher complexities than the SA-GMP. Hence, the proposed SA-GMP has a good convergence performance and much lower computational complexity. In addition, with the increase of the number of users or antennas, more improvement on complexity and performance can be achieved by the SA-GMP. It should be noted that, the MSE of GMP, Jacobi algorithm, AMP, and GaBP algorithm are divergent in this case as the $\beta=10/7$ is very close to 1. Hence, there is no complexity curves for them. However, even when all the methods are convergent for the large $\beta$, \emph{i)} the complexity of GMP is also larger than the SA-GMP due to its lower convergence speed than the SA-GMP, and \emph{ii)} the complexities of Jacobi and GaBP algorithm are larger than the Richardson method due to the lower convergence speed.

\subsection{SA-GMP with Digital Modulation}
{Regarding the selection of error-correcting code, any capacity-approaching multiuser code (e.g. those that can achieve within 2dB away from Shannon limit) can be used. In this paper, we employ the Turbo Hadamard code as an example, and use the superposition coded modulation (SCM) to achieve a Gaussian-like transmission waveform. We acknowledge that the selection of such multiuser code may not be optimal, hbut since the focus of this paper is on the design of gaussian message passing detection, the multiuser code design is beyond the scope of this paper. For more information on the latter, the readers may refer to \cite{Song2017,Lei20161} on the optimization of multiuser SCM (superposition coded modulation), multiuser LDPC or multiuser IRA codes.}

{Fig. \ref{f8} presents the bit error rate (BER) performances of the SA-GMP and LMMSE detection for a practical $30\times15$ overloaded MU-MIMO transmitting digital modulation waveforms, where $\beta=2$. The original GMP diverges in this case. In this system, each user is encoded with a Turbo Hadamard error-correcting code \cite{Ping2003_2,Ping2004}(3 component codes, Hadamard order: 5, spread length: 4), where the code rate is 0.01452 bits/symbol and the code length is $N_c=2.82\times 10^5$. A 10-bit superposition coded modulation \cite{Ma2004, Gadkari1999} is employed for each user to produce Gaussian-like transmitting signals. Hence, the transmitting length of each user is $N=2.82\times 10^4$, the rate of each user is $R_u=0.1452$ bits/symbol, and the sum rate is 4.356 bits per channel use. $E_b/N_0$ is calculated by $E_b/N_0=\frac{P_u}{2R_u\sigma^2_n}$, where $P_u=1$ is the power of each user, and $\sigma^2_n$ is the variance of the Gaussian noise. The BS recovers the messages of all the users by iterative processing between the detector and separate user decoders. In this simulation, we let $N_{ite}^{ese}=1$, which decreases the complexity of SA-GMP to $\mathcal{O}(N_uN_sN_{ite}^{out})$. Fig. \ref{f8} shows that it only needs about 7 iterations for the proposed SA-GMIPD to converge to the LMMSE detection. Furthermore, we can also see that the proposed theoretical results even work for overloaded MU-MIMO systems with a small number of antennas and users ($N_u=30, N_s=15$).}
\section{Conclusion}
This paper considers the low-complexity GMP for uplink multi-user detection of overloaded coded massive MIMO-NOMA. The convergence of the traditional GMP is analysed, resulting in the proof that its variance always converge to the LMMSE detection variance. Furthermore, the convergence region of GMP is given by $N_u/N_s>(\sqrt{2}-1)^{-2}$, but the converged MSE is higher than that of LMMSE multi-user detector. To overcome this limitation, a novel SA-GMP is proposed by minimizing the spectral radius. The resultant SA-GMP is shown to always converge to the LMMSE detection in both mean and variance for any overloaded massive MIMO-NOMA (i.e. $N_u/N_s>1$), with a faster convergence speed  and no higher computational complexity than the traditional GMP.

\appendices
\section{Proof of Proposition 2}
From the expressions (\ref{np1}) and (\ref{var_mes}) of GMP, we have %\vspace{-0.3cm}
\begin{equation}\label{p1}
\!\!{v_{k \to m}^u(\!\tau\!) \!=\!\! {{\Big( {\sum\limits_{i\ne m}\!\! {h_{ik}^2{{( {\sum\limits_{j } {h_{ij}^2v_{j \to i}^u(\tau\!-\!1) \!+\! \sigma _n^2\;} } )}^{ \!-\! 1}} \!\!\!+\! \bar{v}_k^{l^{-1}}\;} } \!\!\Big)}^{ - 1}}}\!\!\!.%\vspace{-0.2cm}
\end{equation}
{As the initial value $\mathbf{V}_{us}(0)$ is equal to $+\boldsymbol{\infty}$, it is easy to find that $\mathbf{V}_{us}(\tau) > \mathbf{0}${\footnote{All the inequations in this paper are component-wise inequalities, i.e., ``$\mathbf{A}_{N_s\times N_u} > \mathbf{B}_{N_s\times N_u}$" denotes ``$a_{ij} > b_{ij}$, $\forall i\in \mathcal{N}_s, \forall j\in \mathcal{N}_u$".}} for any $\tau>0$ during the iteration. Hence, $\mathbf{V}_{us}(\tau)$ has a lower bound $\mathbf{0}$. From (\ref{p1}), we can see that $\mathbf{V}_{us}(\tau)$ is a monotonically non-increasing function with respect to $\mathbf{V}_{us}(\tau-1)$. Besides, we have $\mathbf{V}_{us}(1)<\mathbf{V}_{us}(0)=+\boldsymbol{\infty}$ in the first iteration. Therefore, it can be shown that $\mathbf{V}_{us}(\tau)\leq \mathbf{V}_{us}(\tau-1)$ with $\mathbf{V}_{us}(1) \leq \mathbf{V}_{us}(0)$ by the monotonicity of the iteration function. This means that $\{\mathbf{V}_{us}(\tau)\}$ monotonically decreased with respect to $\tau$ but is lower bounded. Thus, $\{\mathbf{V}_{us}(\tau)\}$ converges to a certain value, i.e., $\mathop {\lim }\limits_{\tau \to  \infty } {\mathbf{V}_{us}}(\tau)=\mathbf{V}^*$.}

{With the symmetry ($\bar{v}_i^l=\bar{v}^l$ for $\forall i\in \mathcal{N}_u$) of the entries in $\mathbf{V}^*$, we have $v_{k\to m}^*=\hat{v}, k\in \mathcal{N}_u\,$, $m\in \mathcal{N}_s$. Since $\frac{1}{N_s}\sum\limits_{i\ne m}  h_{ik}^2\to 1$ and $\frac{1}{N_u} \sum\limits_{j }  h_{ij}^2\to1$, the asymptotic variance of GMP is given by the SE \cite{Donoho2010a,Donoho2009,Bayati2011} below.
\begin{equation}\label{p2}
{\hat{v} = {{\left( {N_s {{{(N_u \hat{v} + \sigma_n^2 )}^{ - 1}} + \bar{v}^{l^{-1}}} } \right)}^{ - 1}}}.%\vspace{-0.2cm}
\end{equation}
Then $\hat{v}$ is the positive solution of (\ref{p2}), i.e.,%\vspace{-0.2cm}
\begin{equation*}\label{p42}
\hat{v} \!=\!\! \frac{{\sqrt {{{\!(\sigma _n^2\bar{v}^{l^{-1}} \!+ \!N_s\! -\! N_u)}^2}\!\!\! +\! 4N_u \bar{v}^{l^{-1}}\sigma _n^2} \! - \!(\sigma _n^2\bar{v}^{l^{\!-\!1}} \!\!\!+ \!N_s\! - \!N_u)}}{{2N_u \bar{v}^{l^{-1}}}}.%\vspace{-0.2cm}
\end{equation*}
Thus, we have Proposition 2.}

\section{Proof of Lemma 1}
From (\ref{np1}) and (\ref{var_mes}), we have%\vspace{-0.2cm}
\begin{eqnarray}\label{p9}
x_{m\to k}^s(\tau)=y_m-\sum\limits_{i}{h_{mi}v_{i\to m}^u(\tau-1)}\cdot\qquad\qquad\qquad\nonumber\\
\Big[ \sum\limits_{j\ne m}h_{ji}v_{j\to i}^{s^{-1}}(\tau-1)x_{j\to i}^s(\tau-1) + \bar{v}_i^{l^{-1}}\bar{x}_i^l\Big].%\vspace{-0.2cm}
\end{eqnarray}
From (\ref{p8}) and (\ref{p6}), the $v^s_{m\to k}(\tau)$ and $v^u_{k\to m}(\tau) $ converge to $v^s$ and $\hat v$ respectively. Therefore, (\ref{p9}) can be rewritten as\vspace{-0.2cm}
\begin{equation}\label{p10}
x_{m\to k}^s(\tau)=y_m-\sum\limits_{i}{h_{mi}}( \gamma\sum\limits_{j\ne m}h_{ji} x_{j\to i}^s(\tau-1) + \alpha\bar{x}_i^l),\vspace{-0.2cm}
\end{equation}
where $\alpha=\hat v/\bar{v}^l$ and $\gamma=\hat v/{v}^s$. Then, we obtain $x_{ m\to k}^s(\tau)=x^s_{m}(\tau)$, $k \in \mathcal{N}_u$ and $m \in \mathcal{N}_s$. Thus, the above equation is rewritten as\vspace{-0.2cm}
\begin{equation}\label{ne_matr_f}
\mathbf{x}^s(\tau) = \textbf{{y}} -  \gamma (\mathbf{H}\mathbf{H}^T-\mathbf{D}_{\mathbf{H}{\mathbf{H}^T}}) \mathbf{x}^s(\tau-1)-\alpha \mathbf{H}\bar{\mathbf{x}}^l,\vspace{-0.2cm}
\end{equation}
where  $\mathbf{x}^s(\tau)={\left[ {{{x}_1^s}(\tau)\;{{x}_2^s}(\tau)\; \cdots \;{{x}_{N_s}^s}(\tau)} \right]^T}$, $\mathbf{D}_{\mathbf{H}\mathbf{H}^T}=diag\{d_{11},d_{22}, \cdots ,d_{N_sN_s}\}$ is a diagonal matrix, and $\{d_{mm}=\mathbf{h}_m^T\mathbf{h}_m, m\in \mathcal{N}_s\}$ are the diagonal entries of matrix $\mathbf{H}\mathbf{H}^T$. Thus, we obtain Lemma 1 with (\ref{a1}).

\section{Proof of Theorem 2}
As $\gamma= \mathcal{O}(\frac{1}{N_u})$, from LLN, the matrix $\gamma\mathbf{D}_{\mathbf{H}\mathbf{H}^T}$ can be approximated by $\gamma N_u\mathbf{I}_{N_s}$. Assuming that the sequence $\{\mathbf{x}^s(\tau)\}$ converges to $\mathbf{x}^*$, then from Lemma 1, we have\vspace{-0.2cm}
\begin{equation}\label{p12}
{\mathbf{x}^*} = {\left( {(1-\gamma N_u){\mathbf{I}_{N_s}} +\gamma\mathbf{H}{\mathbf{H}^T}} \right)^{ - 1}}(\textbf{{y}}-\alpha \mathbf{H} \bar{\mathbf{x}}^l).\vspace{-0.2cm}
\end{equation}
From (\ref{np3}), the {\emph{a-posteriori} estimation} is\vspace{-0.2cm}
\begin{equation}\label{g_fm1}
{\hat{\mathbf{x}}} = \gamma \mathbf{H}^T\mathbf{x}^*+\alpha \bar{\mathbf{x}}^l.\vspace{-0.2cm}
\end{equation}
It means that the GMP converges to $\hat{\mathbf{x}}$ if it converges. With (\ref{p12}) and (\ref{g_fm1}), we have%\vspace{-0.2cm}
 \begin{eqnarray}\label{g_fm2}
{\hat{\mathbf{x}}}
= \left( \theta \mathbf{H}^T\mathbf{H}+ I_{N_u} \right)^{ - 1}\left( \theta \mathbf{H}^T\textbf{{y}}+ \alpha \bar{\mathbf{x}}^l\right),\vspace{-0.2cm}
\end{eqnarray}
where $\theta=(\gamma^{-1}-N_u)^{-1}=\hat v/ \sigma^2_n$, and the second equation is based on the \emph{matrix inverse lemma}. Let $ \textbf{\emph{c}}= \textbf{{y}}-\alpha \mathbf{H}\bar{\mathbf{x}}^l$ and $\mathbf{B}= -  \gamma (\mathbf{H}\mathbf{H}^T-\mathbf{D}_{\mathbf{H}{\mathbf{H}^T}})$, then (\ref{p11}) is a classical iterative algorithm (\ref{a1}). Thus, we can get Theorem 2 based on Proposition 3.\vspace{-0.3cm}

\section{Proof of Corollary 2}
We approximate the input ESE message as $\bar{\mathbf{x}}^l=\bar{v}^l \sigma_{n^l}^{-2}(\mathbf{x}+\mathbf{n}^l)$ where $\mathbf{x}\sim \mathcal{N}^{N_u}({0,\sigma_x^2})$,  $\mathbf{n}^l\sim \mathcal{N}^{N_u}(0,\sigma_{n^l}^{2})$ and $\bar{v}^l=(\sigma_x^{-2}+\sigma_{n^l}^{-2})^{-1}$. It is easy to verify that $\bar{\mathbf{x}}^l\sim \mathcal{N}^{N_u}(0,\bar{v}^l)$ . Then, the GMP can be rewritten to%\vspace{-0.2cm}
\begin{equation*}
\hat{\mathbf{x}}\!=\!\! {\mathbf{x}} \!+\! \left( \sigma_{n}^{\!-2} \mathbf{H}^T\!\mathbf{H}\!+\! \hat{v}^{\!-1} \!I_{N_u} \right)^{\! - \!1}\!\!\left(\! (\sigma_{n^l}^{\!-2}\!-\!\hat{v}^{\!-1}){\mathbf{x}}\!+\! \sigma_{n^l}^{\!-2}\!\mathbf{n}^l \!+\! \sigma_{n}^{\!-2} \mathbf{H}^T\!\mathbf{n}\!\right).%\vspace{-0.2cm}
\end{equation*}
The MSE of the GMP is given by%\vspace{-0.2cm}
\begin{eqnarray}\label{Large_MSE}
\!\!\!\!\!\!\!\!\!\!\!\!&&MSE_{\small GMP}=\frac{1}{N_u}E\left[\|\hat{\mathbf{x}}-\mathbf{x}\|^2_2\right] \nonumber\\
 \!\!\!\!\!\!\!\!\!\!\!\!&&=\tfrac{1}{N_u}E\left[\|( \sigma_{n}^{\!-\!2} \mathbf{H}^T\!\mathbf{H}\!+\! \hat{v}^{\!-1}\! I_{N_u}\! )^{ \!- 1}( \alpha{\mathbf{x}}\!+\! \sigma_{n^l}^{\!-2}\mathbf{n}^l \!+\! \sigma_{n}^{\!-2} \mathbf{H}^T\mathbf{n})\|^2_2\right]\nonumber\\
\!\!\!\!\!\!\!\!\!\!\!\! &&\mathop  > \limits^{(a)}\tfrac{1}{N_u}\!E\!\left[\|( \sigma_{n}^{\!-2} \mathbf{H}^T\!\mathbf{H}\!+\! \bar{v}^{l^{\!-1}}\! I_{N_u} )^{\! - 1}( \alpha{\mathbf{x}}\!+\! \sigma_{n^l}^{\!-2}\mathbf{n}^l \!+\! \sigma_{n}^{-2} \mathbf{H}^T\mathbf{n})\|^2_2\right]\nonumber\\
 %&=& \frac{1}{N_u}E\left[\|( \sigma^{-2}_n\mathbf{H}^T\mathbf{H}+ \hat{v}^{-1} I_{N_u} )^{ - 1}( \sigma_{x}^{-2}{\mathbf{x}}+ \sigma_{n^l}^{-2}\mathbf{n}^l + \sigma_{n}^{-2} \mathbf{H}^T\mathbf{n})\|^2_2\nonumber\right]\\
 \!\!\!\!\!\!\!\!\!\!\!\!&&\mathop  = \limits^{(b)}  \frac{1}{N_u}\;  {\mathrm{tr}\!\left[( \sigma^{-2}_n\mathbf{H}^T\mathbf{H}+ {\bar{v}^{l^{-1}}} I_{N_u} )^{ - 1}\right]} \nonumber\\
 \!\!\!\!\!\!\!\!\!\!\!\!&&= MSE_{\small mmse},%\vspace{-0.2cm}
\end{eqnarray}
where $\alpha=\sigma_{n^l}^{\!-2}\!-\!\hat{v}^{\!-1}$, equation $(b)$ comes from (\ref{PA5}), and the inequality (a) is derived by $\hat{v}<\bar{v}^l$, $\bar{v}^l=(\sigma_x^{-2}+\sigma_{n^l}^{-2})^{-1}$ and the monotonicity of $MSE_{\small GMP}(\hat{v}^{-1})$ with respect to $\hat{v}^{-1}$. Next, we show the detailed proof of monotonicity of $MSE_{\small GMP}(\hat{v}^{-1})$. Let%\vspace{-0.2cm}
\begin{eqnarray*}
\!\!\!\!\!\!\!\!\!\!\!\!&&\mathbf{F}(\hat{v}^{-1})\!=\! \mathbf{W}_{\hat{v}}^{ \!-\! 1}\!\!\left(\! \sigma_{n}^{-2} \mathbf{H}^T\!\mathbf{H} \!+\!\! \left((\hat{v}^{-1}\!\!-\!\sigma_{n^l}^{-2} )^2 \sigma_x^{-2}\!\!+\! \sigma_{n^l}^{-2} \right)\!I_{N_u}\!\right)\!\mathbf{W}_{\hat{v}}^{ - 1}\!\!,\qquad\\
\!\!\!\!\!\!\!\!\!\!\!\!&&\mathbf{W}_{\hat{v}}=( \sigma_{n}^{-2} \mathbf{H}^T\!\mathbf{H}\!+\! \hat{v}^{-1} \!I_{N_u}\! ),%\vspace{-0.2cm}
\end{eqnarray*}
where $\hat{v}^{-1}> \bar{v}^{l^{-1}}>\sigma_{n^l}^{-2}$ and $\bar{v}^{l^{-1}} = (\sigma_x^{-2} + \sigma_{n^l}^{-2} )$. Then, we have
\begin{eqnarray}\label{partial}
\!\!\!\!\!\!\!\!\!\!\!\!&&\frac{\partial \mathbf{F}(\hat{v}^{-1})}{\partial\hat{v}^{-1}}\nonumber\\
\!\!\!\!\!\!\!\!\!\!\!\!&&= 2 \left((\hat{v}^{-1}\!-\!\sigma_{n^l}^{-2}) \sigma_x^{-2}-1\right) \mathbf{W}_{\hat{v}}^{ - 2}\!\left[ I_{N_u}- \left(\hat{v}^{-1}\!\!-\!\sigma_{n^l}^{-2} \right)\!\mathbf{W}_{\hat{v}}^{ - 1}\right]  \nonumber\\
\!\!\!\!\!\!\!\!\!\!\!\!&&\mathop  > \limits^{(a)} 2\! \left(\!(\hat{v}^{-1}\!-\!\sigma_{n^l}^{-2}) \sigma_x^{\!-\!2}\!-\!1\!\right) \mathbf{W}_{\hat{v}}^{ \!-\! 2}\!\left[ \!I_{N_u}\!\!-\!\! \left(\hat{v}^{-1}\!\!-\!\sigma_{n^l}^{\!-\!2} \right) \! \left(\hat{v}^{\!-\!1}I_{N_u}\right)^{\!-\!1}\right] \nonumber\\
\!\!\!\!\!\!\!\!\!\!\!\!&&= 2 \hat{v}\sigma_{n^l}^{-2} \left((\hat{v}^{-1}\!-\!\sigma_{n^l}^{-2}) \sigma_x^{-2}-1\right) \mathbf{W}_{\hat{v}}^{ - 2} \nonumber\\
\!\!\!\!\!\!\!\!\!\!\!\!&&\mathop  > \limits^{(b)}\mathbf{0},\vspace{-0.2cm}
\end{eqnarray}
where the inequality (a) and inequality (b) are based on $\left((\hat{v}^{-1}\!-\!\sigma_{n^l}^{-2}) \sigma_x^{-2}-1\right)>0$, and $\mathbf{H}^T\!\mathbf{H}$ is positive definite. From (\ref{partial}), we get that $MSE_{\small GMP}(\hat{v}^{-1})=\mathrm{tr}\!\left[\mathbf{F}(\hat{v}^{-1})\right]$ is a monotone increasing function with respect to $\hat{v}^{-1}$. Therefore, we have $MSE_{\small GMP}(\hat{v}^{-1})>MSE_{\small GMP}(\bar{v}^{l^{-1}})$ due to $\hat{v}^{-1}>\bar{v}^{l^{-1}}$. As a result, we have the inequality (a) in (\ref{Large_MSE}). Hence, we obtain Corollary 2.

\section{Proof of Lemma 2}
From (\ref{FC1}) and (\ref{FC2}), we have
\begin{eqnarray}\label{FC6}
x_{m\to k}^s(\!\tau\!)\!=\!y'_m\!\!-\!\!\sum\limits_{i}{h'_{mi}\bar{v}_{i}^l}\Big( \sum\limits_{j\ne m}h'_{ji}v_{j}^{s^{\!-1}}\!x_{j\to i}^s(\!\tau\!-\!1\!) \qquad \!\nonumber\\+ \bar{v}_i^{l^{-1}}\!\bar{x}_i^l\Big)
-(w\!-\!1)x_{m \to k}^s(\tau\!-\!1).\;\;
\end{eqnarray}
From (\ref{FCpara}), (\ref{FC6}) can be rewritten to
\begin{eqnarray}\label{FC7}
x_{m\to k}^s(\!\tau\!)\!=\!y'_m\!\!-\!\!\sum\limits_{i}\!{h'_{mi}}\Big( \tilde{\gamma}\sum\limits_{j\ne m}h'_{ji} x_{j\to i}^s(\tau\!-\!1) + \bar{x}_i^l\Big)\qquad \!\nonumber\\
-(w\!-\!1)x_{m \to k}^s(\tau\!-\!1).%\vspace{-0.2cm}
\end{eqnarray}
Then, we have $x_{m\to k}^s(\tau)=x_{m}^s(\tau)$, $\forall k \in \{ 1, \cdots ,K\}$. Thus, (\ref{FC7}) can be rewritten to\vspace{-0.2cm}
\begin{equation*}\label{FC8}
\mathbf{x}^s(\!\tau\!)\! =\! \textbf{{y}}' \!- \! \left[\!\tilde{\gamma} (\mathbf{H}'\mathbf{H}'^T \!-\!\mathbf{D}_{\mathbf{H}'{\mathbf{H}'^T}}) \!+\!(w\!-\!1)\mathbf{I}_{N_s}\!\right] \mathbf{x}^s(\!\tau\!-\!1\!)-\! \mathbf{H}'\bar{\mathbf{x}}^l.
\end{equation*}
As $\gamma= \mathcal{O}(\frac{1}{N_u})$, from LLN, the matrix $\gamma\mathbf{D}_{\mathbf{H}\mathbf{H}^T}$ can be approximated by $\gamma N_u\mathbf{I}_{N_s}$.  Thus, we get Lemma 2 with (\ref{FC8}).

\section{Proof of Theorem 3}
Let $\mathbf{c}=\textbf{{y}}'-\mathbf{H}'\bar{\mathbf{x}}^l$ and $\mathbf{B}=\tilde{\gamma} \mathbf{H}'\mathbf{H}'^T +\big((1-\tilde{\gamma}N_u )w-1\big)\mathbf{I}_{N_s}$. SA-GMP is then a classical iterative algorithm. Thus, according to \emph{Proposition 3}, the SA-GMP converges if\vspace{-0.2cm}
\begin{equation}\label{FC9}
\rho\left(\tilde{\gamma} \mathbf{H}'\mathbf{H}'^T +\big((1-\tilde{\gamma}N_u )w-1\big)\mathbf{I}_{N_s}\right)<1, \vspace{-0.2cm}
\end{equation}
i.e., $\rho\left(\mathbf{I_{N_s}}-w\mathbf{A}\right)<1$ and $\mathbf{A}= \tilde{\gamma}\mathbf{H}\mathbf{H}^T +(1-\tilde{\gamma}N_u )\mathbf{I}_r$.
When $N_u$ and $ N_s$ are large, the smallest and largest eigenvalues ${\lambda}_{min}^\mathbf{A}$ and ${\lambda}_{max}^\mathbf{A}$ of matrix $\mathbf{A}$ are given by \cite{verdu2004}%\vspace{-0.2cm}
\BS\label{eigv1}\begin{eqnarray}
&&{\lambda}_{min}^\mathbf{A}\!=\!1+\tilde{\gamma} N_u{(( {1 \!- \!\sqrt {\beta^{-1}} } )^2\!\!-\!1)},\\
&&{\lambda}_{max}^\mathbf{A}\!=\!1+\tilde{\gamma} N_u{(( {1 \!+ \sqrt {\beta^{-1}} } )^2\!-\!1)}.%\vspace{-0.2cm}
\end{eqnarray}\ES
From (\ref{eigv1}) and $\tilde{\gamma}=(N_u+snr^{-1})^{-1}<N_u^{-1}$, we have ${\lambda}_{max}^\mathbf{A}>{\lambda}_{min}^\mathbf{A}>0$, which means that $\mathbf{A}$ is strictly positive definite. Therefore, the condition (\ref{FC9}) can always be satisfied if $0<w<2/\lambda_{max}^\mathbf{A}$. In addition, we can always find such a $w$ that satisfies $0<w<2/\lambda_{max}^\mathbf{A}$ because ${\lambda}_{max}^\mathbf{A}>0$.

Next, we will show that the SA-GMP converges to the LMMSE detection. From {Proposition 3}, we know that the sequence $\{\mathbf{x}^s(\tau)\}$ converges to\vspace{-0.2cm}
\begin{equation}\label{FC13}
{\mathbf{x}^*} = \sqrt{w^{-1}}{\left( {(1-\tilde{\gamma} N_u){\mathbf{I}_{N_s}} + \tilde{\gamma}{\mathbf{H}}\mathbf{H}^T} \right)^{ - 1}}(\textbf{{y}}-\mathbf{H}\bar{\mathbf{x}}^l).\vspace{-0.2cm}
\end{equation}
Therefore, from (\ref{FCpara}), the sequence $\{\mathbf{x}^u(\tau)\}$ converges to\vspace{-0.2cm}
\begin{eqnarray*}\label{sa_cv}
{\hat{\mathbf{x}}}\!\!\!\! &=&\!\!\!\! \tilde{\gamma} \mathbf{H}'^T\mathbf{x}^* + \bar{\mathbf{x}}^l\\
\!\!\!\! &=&\!\!\!\! \tilde{\gamma} \mathbf{H}^T{\left( {(1-\tilde{\gamma} N_u){\mathbf{I}_{N_s}} + \tilde{\gamma}{\mathbf{H}}\mathbf{H}^T} \right)^{ - 1}}(\textbf{{y}}-\mathbf{H}\bar{\mathbf{x}}^l)+\bar{\mathbf{x}}^l.\vspace{-0.2cm}
\end{eqnarray*}
Similar to (\ref{g_fm2}), we apply the \emph{Matrix Inversion Lemma} to (\ref{sa_cv}), and get\vspace{-0.2cm}
\begin{equation}
{\hat{\mathbf{x}}} = \left(  snr \mathbf{H}^T\mathbf{H}+ I_{N_u} \right)^{ - 1}\left( snr \mathbf{H}^T\textbf{{y}} + \bar{\mathbf{x}}^l\right),\vspace{-0.25cm}
\end{equation}
which is the same as the LMMSE detection (\ref{f_gm}). Therefore, we get the Theorem 3.

\section{Proof of Corollary 3}
 The relaxation parameter $w$ can be optimized by \cite{Gao2014}\vspace{-0.2cm}
\begin{equation}\label{FC10}
w^* =2/({\lambda}_{min}^\mathbf{A}+{{\lambda}_{max}^\mathbf{A}}).\vspace{-0.2cm}
\end{equation}
It minimizes the spectral radius of $\mathbf{I}_{N_s}-w\mathbf{A}$ to $\rho_{min}(\mathbf{I}_{N_s}-w\mathbf{A})= \frac{{\lambda}_{max}^\mathbf{A}-{\lambda}_{min}^\mathbf{A}}{{\lambda}_{max}^\mathbf{A}+{\lambda}_{min}^\mathbf{A}}<1$ . From (\ref{eigv1}), we obtain the optimal relaxation parameter\vspace{-0.2cm}
  \begin{equation}\label{FC_rev1}
 w^* = {1 \mathord{\left/
 {\vphantom {1 {\left( {1 + \tilde{\gamma} M \beta} \right)}}} \right.
 \kern-\nulldelimiterspace} {\left( {1 + \tilde{\gamma} N_u \beta^{-1}} \right)}}=1/(1+\tilde{\gamma} N_s),\vspace{-0.2cm}
 \end{equation}
 and the minimal spectral radius in SA-GMP\vspace{-0.2cm}
 \begin{equation}\label{FC12}
 \rho_{_{SA-GMP}} = \rho_{min}(\mathbf{I}_{K}-w\mathbf{A})=\frac{2\tilde{\gamma}\sqrt{N_sN_s}}{1+\tilde{\gamma} N_s}<1,%\vspace{-0.2cm}
 \end{equation}
 where $\tilde{\gamma}={({N_u} + sn{r^{ - 1}})^{ - 1}}$ according to (\ref{FC4}). Comparing (\ref{FC12}) with (\ref{radius1}), we have%\vspace{-0.2cm}
 \begin{equation}\label{FC_rev2}
 \frac{\rho_{_{GMP}}}{\rho_{_{SA-GMP}}} = (1+\tilde{\gamma} N_s)^2>1.%\vspace{-0.2cm}
 \end{equation}
With (\ref{FC_rev1})$-$(\ref{FC_rev2}), we have Corollary 3.

\ifCLASSOPTIONcaptionsoff
  \newpage
\fi

\end{document}